\documentclass[aip, jmp, amsmath,amssymb,]{revtex4}

\usepackage{graphicx}
\usepackage{dcolumn}
\usepackage{bm}

\usepackage[utf8]{inputenc}
\usepackage[T1]{fontenc}
\usepackage{mathptmx}
\usepackage{etoolbox}
\usepackage{enumerate}
\usepackage{amsthm}
\usepackage{color}
\usepackage{graphicx}
\usepackage{epstopdf}
\usepackage{float}
\usepackage{thmtools}
\usepackage{thm-restate}
\usepackage{hyperref}
\usepackage{cleveref}
\usepackage{tabularx}
\usepackage{tikz,amstext}

\newtheorem{thm}{Theorem}[section]

\newtheorem{prop}[thm]{Proposition}

\newtheorem{notation}{Notation}[section]
\newtheorem{assumption}{Assumption}[section]
\newtheorem{defn}{Definition}[section]
\newtheorem{rem}{Remark}[section]
\newtheorem{example}{Example}[section]

\newcommand*\myrect[1]{%
  \begin{tikzpicture}
    \node[draw,rectangle,inner sep=0pt] {#1};
  \end{tikzpicture}}
\newcommand{\un}{1\mkern -4mu{\rm l}}
\newcommand{\A}{\mathcal{A}}

\newcommand{\R}{\mathbb{R}}

\newcommand{\sqsum}{\mathbin{\text{\myrect{\textnormal{+}}}}}

\DeclareMathOperator{\tr}{tr}

\makeatletter
\def\@email#1#2{%
 \endgroup
 \patchcmd{\titleblock@produce}
  {\frontmatter@RRAPformat}
  {\frontmatter@RRAPformat{\produce@RRAP{*#1\href{mailto:#2}{#2}}}\frontmatter@RRAPformat}
  {}{}
}%
\makeatother
\begin{document}

\preprint{AIP/123-QED}

\title[The spectrum of local random Hamiltonians]{The spectrum of local random Hamiltonians}
\author{B. Collins}
\altaffiliation{Department of Mathematics, Graduate School of Science, Kyoto University, Japan}
\author{Z. Yin}
\altaffiliation{School of Mathematics and Statistics, Central South University, China}%
\author{L. Zhao}
\altaffiliation{School of Mathematics, Harbin Institute of Technology, China}
\author{P. Zhong}
\altaffiliation{Department of Mathematics and Statistics, University of Wyoming, USA}

\date{\today}

\begin{abstract}
The spectrum of a local random Hamiltonian can be represented generically by the so-called $\epsilon$-free convolution of its local terms' probability distributions. We establish an isomorphism between the set of $\epsilon$-noncrossing partitions and permutations to study its spectrum. Moreover, we derive some lower and upper bounds for the largest eigenvalue of the Hamiltonian. 
\end{abstract}

\maketitle


\section{Introduction}

In general, the Hamiltonian of a many-body system is local; namely, the Hamiltonian can be derived from a sum of local terms which describe the interactions between the local systems. For instance, if the particles of a quantum spin only correlate with the short-range, then the total Hamiltonian of the spin is local. Understanding their Hamiltonians' spectrum is essential for their theoretical and experimental aspects \cite{AFOV08, M17, BAL19}. Furthermore, if the interactions are disordered, many striking phenomena appear. For example, the upper bound of the speed of information propagation through a quantum spin chain with disordered interactions may be significantly lower than the famous Lieb-Robinson's bound \cite{LR72}, which indicates the system exhibits the phenomenon of Anderson localization \cite{BO07}.

In this paper, we consider the following general model: let $n$ be the number of local systems and let $d$ be the local dimension. Thus,  the Hilbert space, which describes the total system, is given by an $n$-fold tensor product $(\mathbb{C}^d)^{\otimes n}.$ Let $\mathcal{K}_\iota, \iota=1, \ldots, m$ be subsets of the integer set $\{1, \ldots,n\}$ and suppose that the interactions only occur between the sites in $\mathcal{K}_\iota.$
As a result, the system's total Hamiltonian $H$ of the system is given by the sum of local terms, namely, 
\begin{equation}
H = \sum_{\iota=1}^{m} H_\iota,
\end{equation}
where $H_\iota \in \bigotimes_{s \in \mathcal{K}_\iota} \mathbb{M}_d(\mathbb{C}) \otimes \bigotimes_{s \notin \mathcal{K}_\iota}(\mathbb{C}\un_d).$ To express the disorderliness of the interactions, we can ideally assume that $H_\iota$'s are sampled by some given random ensembles. The above general framework covers a lot of explicit models, such as the Heisenberg, Ising, XY, XXZ, AKLT models, etc., which are commonly studied in quantum many-body physics. Thus, a natural question arises: can one figure out the spectrum of given local random Hamiltonians? Unfortunately, obtaining an analytical result is not easy since, generally, the local terms do not commute. For quantum spin models ($d=2$), the Jordan-Wigner transformation can be used to reduce the problem to an equivalent single Hamiltonian \cite{HSS12}. However, a large amount of work relies on numerical analysis.

Another line of investigation is carried out in the asymptotic picture ($d \rightarrow \infty$). One can consider the random Hamiltonian $H$ in the framework of random matrix theory and study its spectrum in the large $d$ limit. 
The idea may date back to the 1950s when Wigner studied a model of complicated nuclei. The Hamiltonian of the system can be equivalently considered as a large dimensional random matrix, and the distribution of its eigenvalues converges to the celebrated Wigner's semicircular law \cite{Wig55}. Later, when studying the $O(d)$ and $U(d)$ quantum field theories, 't Hooft discovered that remarkable simplifications occur in the large $d$ limit \cite{Hooft, Zee}. Occasionally but surprisingly, the abovementioned theory satisfies Voiculescu's free probability theory \cite{D94}. Indeed, the limit law as $d\rightarrow \infty$ can be described using random variables in the framework of free probability \cite{BIPZ78, FZ97, G2010, CM16}.

In \cite{ML19}, Morampudi and Laumann proposed a systematic pictorial method to calculate the correlators based on randomly interacting spin systems with spatial locality.  According to 't Hooft's idea, only the so-called monochromatic stacked planar diagrams survive in the large $d$ limit. Consequently, in the large  $d$ limit, an algebraic relation known as "heap-freeness" emerges among $H_\iota$'s. This phenomenon
was also independently observed by Charlesworth and Collins \cite{CC2015}. Due to their theory, heap-freeness is equivalently called $\epsilon$-freely independence, a mixture of classical and free independence \cite{M04, SW16}. Here $\epsilon$ is a symmetric matrix given by $\mathcal{K}_\iota$'s in our setting. 

Therefore, motivated by the free probability theory, we show that the limit of the spectrum of 
$H$ can be given by the so-called $\epsilon$-free convolution (see Definition \ref{def:convolution}) of distributions of local terms for some given random ensembles (see Theorem \ref{thm:GUE} and \ref{thm:U}). We remark that this conclusion was also implicitly indicated in \cite{ML19}. 
The key point is to establish an isomorphism between the set of all $\epsilon$-noncrossing partitions \cite{M04, SW16, EPS18} (called monochromatic dependency partitions in \cite{ML19}) and the set of all 
$\epsilon$-noncrossing permutations. Moreover, the M\"{o}bius functions of the corresponding set of partitions should also be considered (see Definition \ref{def:permutation} and Theorem \ref{thm:iso}). 
Our second result is that the convergence of the spectrum distribution holds almost surely.
To this end, we study the fluctuation of the correlators and obtain a stronger result of the concentration of measure than the one in \cite{ML19} (see Proposition \ref{prop:almost-GUE} and \ref{prop:almost-U}). Finally, by some rough enumerations of $\epsilon$-noncrossing partitions, we derive some bounds for the largest eigenvalue of $H$ (see Proposition \ref{prop:lowerH-GUE} and \ref{prop:XY-model}). 


\section{Preliminaries}

\subsection{$\epsilon$-noncrossing partition and permutation}

In this paper, we introduce a symmetric matrix $\epsilon$, reflecting the particles' interactions, 
 given by the following definition. We use $[n]$ to denote the integer set $\{1, \ldots, n\}.$ 
\begin{defn}
Let $\mathcal{K}_\iota, \iota=1, \ldots, m$ be subsets of the integer set $[n],$  we define $\epsilon = (\epsilon_{i,j})_{i,j=1}^m$ by a $m \times m$ symmetric matrix with entries $\epsilon_{i,j}$ satisfies
\begin{equation}\label{eq:epsilon}
\epsilon_{i,j}= \begin{cases} 1 & \text{if}\;\; \mathcal{K}_i \cap \mathcal{K}_j =\emptyset;\\
0 & \text{otherwise}.
\end{cases}
\end{equation}
\end{defn}

We will use a bunch of notions in combinatorics and refer to Appendix \ref{sec:preli} for more details. Let $\mathcal{P}(k)$ be the set of all partitions of $[k],$ and let ${\boldsymbol \iota}: [k]\rightarrow [m]$ be a map given by 
$${\boldsymbol \iota} =(\iota_1, \iota_2, \ldots, \iota_k), \iota_i \in [m], i=1, \ldots, k.$$ 
We denote $\ker{{\boldsymbol \iota}}$ by a partition in $\mathcal{P}(k)$ such that $i, j$ belongs to the same block of  $\ker{{\boldsymbol \iota}}$ whenever $\iota_i = \iota_j.$

Thus, for the symmetric matrix $\epsilon$ given in \eqref{eq:epsilon} and a given map ${\boldsymbol \iota}: [k]\rightarrow [m],$ one can define the $(\epsilon, {\boldsymbol \iota})$-noncrossing partitions as follows:
\begin{defn}\cite{SW16, EPS18}.
 A partition $\alpha \in \mathcal{P}(k)$ is called $(\epsilon, {\boldsymbol \iota})$-noncrossing if there exist
$1\leq i < p <j<q \leq k$ such that $i \sim_\alpha j\nsim_\alpha p\sim_\alpha q,$ then we must have $\epsilon_{\iota_i,\iota_p}=1.$ Note that $i \sim_\alpha j$ means the elements $i, j$ are in the same block of $\alpha.$ We define 
$$\mathcal{NC}^{(\epsilon, {\boldsymbol \iota})}(k): = \{ \alpha \in \mathcal{P}(k): \alpha \leq \ker{\boldsymbol \iota} \; \text{and is} \; (\epsilon, {\boldsymbol \iota})\text{-noncrossing}\}.$$ 
It is known that $\mathcal{NC}^{(\epsilon, {\boldsymbol \iota})}(k)$ is a lattice, and the poset order is defined by the refinement of partitions. 
\end{defn}
Denote $\mathcal{NC}(k)$ by the set of all noncrossing partitions of $[k],$ here are some extreme cases. 
\begin{enumerate}[{\rm (i)}]
\item If $\mathcal{K}_\iota$'s are pairwise disjoint, then every non-diagonal entry of $\epsilon$ is $1.$ Then $\mathcal{NC}^{(\epsilon, {\boldsymbol \iota})}(k)= \{ \alpha \in \mathcal{P}(k): \alpha \leq \ker{\boldsymbol \iota} \}.$ 
\item If $\mathcal{K}_1= \mathcal{K}_2=\cdots= \mathcal{K}_m$,  then every entry of $\epsilon$ is $0.$ Then $\mathcal{NC}^{(\epsilon, {\boldsymbol \iota})}(k)= \{ \alpha \in \mathcal{NC}(k): \alpha \leq \ker{\boldsymbol \iota} \}.$ 
\item If $\ker{{\boldsymbol \iota}} \in \mathcal{NC}(k),$ then $\mathcal{NC}^{(\epsilon, {\boldsymbol \iota})}(k)= \{ \alpha \in \mathcal{NC}(k): \alpha \leq \ker{\boldsymbol \iota} \}.$ 
\end{enumerate}

\begin{example}\label{example:1}
Let $m =4$, and take the entries of $\epsilon$ as follows: $\epsilon_{1, 3} = \epsilon_{1, 4} = \epsilon_{2, 4} =1,$ and other entries are $0$'s. 
\begin{enumerate}[{\rm (i)}]
\item Let $k=8$ and ${\boldsymbol \iota}= (\iota_1, \iota_2, \ldots, \iota_8) = (1, 3, 3, 1, 3, 2, 4, 2),$ then $\ker{\boldsymbol \iota}= \{ \{1, 4\}, \{2, 3, 5\}, \{6, 8\}, \{7\}\};$
\item Let $\alpha= \{ \{1, 4\}, \{2\},  \{3, 5\}, \{6, 8\}, \{7\}\},$ then $\alpha \in \mathcal{NC}^{(\epsilon, {\boldsymbol \iota})}(8)$ for the above given ${\boldsymbol \iota}.$ 
\end{enumerate}
\end{example}

We denote the set of permutations on $n$ elements by $S_n$. For a permutation $\pi \in S_n$, we denote  $|\pi|$ by
the minimal number of transpositions needed to decompose $\pi.$ Let $\gamma_n \in S_n$ be the full cycle $\gamma_n = (1,2, \ldots, n).$ For any $\pi \in S_n$, we always have 
$$|\pi|+ |\pi^{-1} \gamma_n| \leq n-1.$$
We call $\pi$ noncrossing if the equality holds, or equivalently we can say that $\pi$ lies on the geodesic path $1_n-\pi-\gamma_n$ in $S_n$. 

\begin{notation}
For given $\epsilon$ and ${\boldsymbol{\iota}}: [k] \rightarrow [m],$ we use the following notations introduced in \cite{CC2015}:
\begin{enumerate}[{\rm (i)}]
\item $S^{({\boldsymbol\iota})}_k$ is the group of permutations $\sigma$ with $\iota_j = \iota_{\sigma(j)}$ for all $j.$ Hence each $\sigma \in S^{({\boldsymbol\iota})}_k$ stabilize $\ker{\boldsymbol \iota}.$ In particular, we denote $\mathcal{P}_2^{(\boldsymbol \iota)}(k)$ by the set of pair partitions of $[k]$ which stabilize $\ker{\boldsymbol \iota}.$
\item For each $s \in [n],$ we denote $J_s:= \{j\in [k]: s \in \mathcal{K}_{\iota_j}\},$ and moreover we denote $k_s: = \# J_s.$ We denote $\gamma_s$ by the full cycle of $J_s$ and by $1_s$ the identity of permutation on $J_s.$
\item Any $\sigma \in S^{({\boldsymbol{\iota}})}_k$ induces a permutation $\sigma_s \in S_{J_s} \simeq S_{k_s},$ since it preserves each $J_s.$
\item For any $\alpha \in \mathcal{P}(k)$, we denote $\alpha_s := \alpha|_{J_s}$ by the restriction of $\alpha$ on $J_s$, and by $\alpha_B: = \alpha|_B$ for any block of $\ker{\boldsymbol \iota}.$
\item We use $\#_{k_s}(\cdot)$ to stress the permutation in the bracket is viewed as an element of $S_{k_s}$ and not as the induced permutation in $S_k$ which is constant on $J_s.$
\end{enumerate}
\end{notation}

Note that $J_s$ is the union of some blocks of $\ker{\boldsymbol \iota}$ and it might be an empty set for some $s \in [n]$ which depends on the map ${\boldsymbol \iota}.$ It is clear that
\begin{equation}
\sum_{s=1}^n k_s = \sum_{\ell=1}^k \# (\mathcal{K}_{\iota_\ell}),
\end{equation}
where $\#(\cdot)$ means the cardinality of a given set. Now we are ready to introduce the notion of $(\epsilon, {\boldsymbol \iota})$-noncrossing permutations.
\begin{defn}\label{def:permutation}
For given $\epsilon$ and ${\boldsymbol{\iota}}: [k] \rightarrow [m],$ 
\begin{enumerate}[{\rm (i)}]
\item We define the set of all  $(\epsilon, {\boldsymbol \iota})$-noncrossing permutations of $[k]$, denoted by $S^{(\epsilon,{\boldsymbol{\iota}})}_{NC}(\gamma_k)$, as follows:
\begin{equation*}
S^{(\epsilon,{\boldsymbol{\iota}})}_{NC}(\gamma_k): =  \{\sigma \in S^{({\boldsymbol{\iota}})}_k : 1_s-\sigma_s-\gamma_s \; \text{for all} \; s \in [n]\}.
\end{equation*}
In other words, the restriction to any $J_s$, $\sigma_s$ is a noncrossing permutation. 
\item For $\sigma, \tau \in  S^{(\epsilon,{\boldsymbol{\iota}})}_{NC}(\gamma_k)$ we say that $\sigma \leq \tau$, if for any $s \in [n]$, $\sigma_s$ and $\tau_s$ lie on the same geodesic and if $\sigma_s$ comes before $\tau_s,$ i.e.,
$$1_s-\sigma_s-\tau_s-\gamma_s, \;\; \text{for all} \; s \in [n].$$ 
\end{enumerate}
\end{defn} 

\begin{example}\label{example:2}
Let $m=4$ and $n=5,$ let $\mathcal{K}_\iota= \{\iota, \iota+1\}, \iota=1, \ldots, 4,$ therefore $\epsilon_{1,3} = \epsilon_{1,4} = \epsilon_{2,4} =1$ and other entries of $\epsilon$ are $0$'s. Moreover, 
let $k=8$ and $\boldsymbol{\iota} = \{\iota_1, \iota_2, \ldots, \iota_8\} = (1, 3, 3, 1, 3, 2, 4, 2).$ Then $S^{({\boldsymbol\iota})}_8$ is generated by the transpositions $(1,4), (6,8), (2,3)$ and $(2,5).$ So that we have 
$$J_1 = \{1,4\}, J_2=\{1,4,6,8\}, J_3 = \{2,3,5,6,8\}, J_4= \{2,3,5,7\}, J_5= \{7\}.$$
If $\sigma = (1, 4) (3, 5) (6,8) \in S_8$ we have 
$$\sigma_1 = (1,4), \sigma_2= (1,4) (6,8), \sigma_3 =(3,5) (6,8), \sigma_4 = (3,5), \sigma_5= \un_{J_5}.$$
It is easy to check that $\sigma \in S^{(\epsilon,{\boldsymbol{\iota}})}_{NC}(\gamma_8)$ for given $\epsilon$ and $\boldsymbol{\iota}.$
\end{example}

In \cite{B97}, Biane showed that there is an isomorphism (of posets) between noncrossing partitions and permutations, which plays an essential role in the aspect of combinatoric free probability theory. Likewise, we can obtain the following 
isomorphism between $\epsilon$-noncrossing partitions and permutations. We refer to Example \ref{example:1} and \ref{example:2} for the isomorphism.

\begin{thm}\label{thm:iso}
With the relation "$\leq$", $S^{(\epsilon,{\boldsymbol{\iota}})}_{NC}(\gamma_k)$ becomes a poset. Moreover, there is a bijection between $ \mathcal{NC}^{(\epsilon, {\boldsymbol \iota})}(k)$ and $S^{(\epsilon,{\boldsymbol{\iota}})}_{NC}(\gamma_k)$ which preserves the poset structure. 
\end{thm}

\begin{proof}
It is elementary to check that $S^{(\epsilon)}_{NC}(\gamma_k)$ is a poset with the given relation.
Let $\alpha$ be a partition of $[k]$; we will denote $P_\alpha$ by the permutation $\pi \in S_k$, which is determined by the following properties:
\begin{enumerate}[(a)]
\item $\alpha$ is $\pi$-invariant, i.e., $\pi$ stabilize the blocks of $\alpha;$
\item if $B= \{i_1, i_2, \ldots, i_s\}$ is a block of $\alpha,$ with $1\leq i_1<i_2<\cdots<i_s\leq k,$ then we have $\pi(i_1)=i_2, \ldots, \pi(i_{s-1}) = i_s, \pi(i_s)=i_1.$	
\end{enumerate}
It induces a map $P: \mathcal{P}(k) \rightarrow S_k$ given by $P(\alpha): = P_\alpha=\pi$ for $\alpha \in \mathcal{P}(k).$ It is known \cite{B97, NS2006} that the map $P$ is a bijection between $\mathcal{NC}(k)$ and $S_{NC}(\gamma_k)$ which preserves the poset structure. 

\noindent{\bf Claim.} $P$ is also an isomorphism between $ \mathcal{NC}^{(\epsilon, {\boldsymbol \iota})}(k)$ and the set $S^{(\epsilon,{\boldsymbol{\iota}})}_{NC}(\gamma_k).$

Firstly, by the definition of $P$, it is clear that for any $\alpha \leq \ker{\boldsymbol \iota}$, $P_\alpha \in S_k^{({\boldsymbol \iota})}.$ 
Now for any $\alpha \in \mathcal{NC}^{(\epsilon, {\boldsymbol \iota})}(k),$ we will show that $\pi_s$ is noncrossing on $J_s$ for every $s \in [n],$ where $\pi = P_\alpha.$ Hence $P(\alpha)\in S^{(\epsilon,{\boldsymbol{\iota}})}_{NC}(\gamma_k).$ Suppose this is not ture, i.e., there exists a $s \in [n]$ such that $\pi_s$ is crossing on $J_s.$ Accordingly there exists $1\leq i<p<j <q \leq k$ such that
\begin{enumerate}[\rm (i)]
\item $i,p,j,q \in J_s;$
\item $i \sim_{B_{1,s}} j, p \sim_{B_{2,s}} q,$ where $B_{1,s}$ and $B_{2,s}$ are two different blocks of $\alpha_s.$ 
\end{enumerate}
By (i), we have $s \in \mathcal{K}_{\iota_i}$ and $s \in \mathcal{K}_{\iota_p}.$ Therefore $\mathcal{K}_{\iota_i} \cap \mathcal{K}_{\iota_p} \neq \emptyset.$ By the definition of $\epsilon$ we can deduce that $\epsilon_{\iota_i, \iota_p}= 0.$ However, by (ii), there must have two different blocks $B_1$ and $B_2$ of $\alpha$ such that  $i \sim_{B_{1}} j, p \sim_{B_{2}} q.$ In fact $B_{i,s} = B_i |_{J_s}.$ Thus 
$\iota_i = \iota_j$ and $\iota_p= \iota_q$. Since $ \alpha \in \mathcal{NC}^{(\epsilon, {\boldsymbol \iota})}(k),$ one has $\epsilon_{\iota_i, \iota_p}=1,$ which is a contradiction.

Conversely, suppose that $\alpha \notin  \mathcal{NC}^{(\epsilon, {\boldsymbol \iota})}(k),$ then there exist a crossing $1\leq i< p< j <q \leq k, i \sim_\alpha j\nsim_\alpha p \sim_\alpha q$ with $\epsilon_{\iota_i, \iota_p}=0.$ By the definition of $\epsilon$ we have $\mathcal{K}_{\iota_i} \cap \mathcal{K}_{\iota_p} \neq \emptyset.$ Hence there exists a $s\in [n]$ such that $i, p \in J_s.$ Moreover, recall that $\iota_i = \iota_j$ and $\iota_p = \iota_q$, we have $i, p, j, q\in J_s.$ So, finally, we find a crossing partition in $J_s$. Thus the restriction of $P_\alpha$ on $J_s$ is crossing, which induces that
$P_\alpha \notin S^{(\epsilon,{\boldsymbol{\iota}})}_{NC}(\gamma_k).$   

Now we turn to show that $P$ preserves the poset structure. Let $\alpha, \beta \in \mathcal{NC}^{(\epsilon, {\boldsymbol \iota})}(k)$ and denote $\pi = P_\alpha, \sigma = P_\beta.$ Suppose that $\alpha \leq \beta.$
Obviously, we have $\alpha_s \leq \beta_s$ for any $s \in [n].$ Now consider the restriction of $P$ on $\mathcal{P}(k_s),$ denote by $P_s.$ Certainly, we have
$P_s$ maps $\alpha_s$ (resp. $\beta_s$) to $\pi_s$ (resp. $\sigma_s$). Hence by Biane's isomorphism \cite{B97, NS2006}, $P_s$ is a bijection between $\mathcal{NC}(k_s)$ and $S_{NC}(\gamma_s)$ which preserves the poset structure. It induces that $1_s-\pi_s-\sigma_s-\gamma_s$ for any $s\in [n],$ which implies $\pi \leq \sigma.$
\end{proof}

\begin{notation} 
For given ${\boldsymbol{\iota}}: [k] \rightarrow m,$ we denote $\mathcal{NC}^{(\epsilon, {\boldsymbol \iota})}_2(k)$ by the set of all pair partitions of $\mathcal{NC}^{(\epsilon, {\boldsymbol \iota})}(k).$
\end{notation}
Since each pair partition in $\mathcal{P}_2(k)$ is identified with a transposition of $S_k$ via the map $P$, then we have 
$$\mathcal{P}_2^{({\boldsymbol \iota})}(k) = \{\sigma \in S_k^{({\boldsymbol \iota})}: \sigma \;\text{is a transposition}\}.$$
Moreover, by the above proposition, we have 
\begin{equation}\label{eq:pair-trans}
\mathcal{NC}^{(\epsilon, {\boldsymbol \iota})}_2(k)= \{\alpha \in \mathcal{P}_2^{({\boldsymbol \iota})}(k) : 1_s -\sigma_s-\gamma_s \; \text{for all} \; s\in [n]\},
\end{equation}
where $\sigma= P_\alpha.$

Let us remark that for the above-mentioned extreme cases (ii) and (iii), one can reduce to Biane's isomorphism between the noncrossing partitions and permutations \cite{B97} (see Proposition \ref{prop:Biane}).


\subsection{Convolution operations and M\"{o}bius inversion}\label{subsec:Mobius-inversion}

Let $\mathcal{P}$ be a finite poset, denote
$$
\mathcal{P}^{(2)}: = \{(\alpha, \beta): \alpha, \beta \in \mathcal{P}, \alpha \leq \beta\}.
$$
For $F, G: \mathcal{P}^{(2)}  \rightarrow \mathbb{C}$, their convolution $F \ast G$ is given by:
$$(F \ast G) (\alpha, \beta) := \sum_{\substack{ \eta \in \mathcal{P},\\ \alpha\leq \eta \leq \beta}} F(\alpha, \eta) G(\eta, \beta).$$
The zeta function $\zeta: \mathcal{P}^{(2)} \rightarrow \mathbb{C}$ is defined by
$$\zeta (\alpha, \beta)=1, \;\; \forall (\alpha, \beta) \in \mathcal{P}^{(2)}.$$
And the unit $\delta: \mathcal{P}^{(2)} \rightarrow \mathbb{C}$   of the convolution is given by
\begin{equation*}
\delta(\alpha, \beta)= \begin{cases} 1 & \text{if}\;\; \alpha = \beta;\\
0 &  \text{if}\;\; \alpha < \beta.
\end{cases}
\end{equation*}
\begin{defn}\cite{NS2006}.
For given poset $\mathcal{P},$ the M\"{o}bius function of $\mathcal{P}$, denoted it by $\mu$, is the inverse of $\zeta$ under convolution, i.e., $\mu \ast \zeta = \delta.$ Equivalently, the M\"{o}bius function is uniquely determined by the following equations
\begin{equation}\label{eq:Mobius-def}
\sum_{\substack{\eta \in \mathcal{P},\\ \alpha \leq \eta \leq \beta}} \mu (\eta, \beta)= \begin{cases} 1 & \text{if}\;\; \alpha = \beta;\\
0 &  \text{if}\;\; \alpha < \beta.
\end{cases}
\end{equation}
To avoid confusion, let us denote $\mu_{\mathcal{P}}(\alpha, \beta)$ by the  M\"{o}bius function of the given poset $\mathcal{P}.$
\end{defn}

\noindent{\it M\"{o}bius function of $\mathcal{NC}^{(\epsilon, {\boldsymbol \iota})}(k)$}-Note that the M\"{o}bius function of $\mathcal{NC}(k)$ can be explicitly computed by the products of Catalan numbers \cite{NS2006}. Here we will give a direct method to compute the M\"{o}bius function of  $\mathcal{NC}^{(\epsilon, {\boldsymbol \iota})}(k)$ by using Theorem \ref{thm:iso}. 

\begin{prop}\label{prop:mobius}
Fix $\epsilon$ and ${\boldsymbol \iota}: [k] \rightarrow [m],$ the  M\"{o}bius function of $\mathcal{NC}^{(\epsilon, {\boldsymbol \iota})}(k)$ can be computed as follows:
\begin{equation}
\mu_{\mathcal{NC}^{(\epsilon, {\boldsymbol \iota})}(k)} (\alpha, \beta) = \prod_{B \in \ker{\boldsymbol \iota}} \mu_{\mathcal{NC}(\#(B))} (\alpha_B, \beta_B),
\end{equation}
for all $\alpha, \beta \in \mathcal{NC}^{(\epsilon, {\boldsymbol \iota})}(k).$
\end{prop}

\begin{proof}
It sufficies to check Equation \eqref{eq:Mobius-def} for the poset $\mathcal{NC}^{(\epsilon, {\boldsymbol \iota})}(k).$ Firstly, suppose that $\alpha = \beta$, thus $\alpha_B = \beta_B$ for every $B \in \ker{\boldsymbol \iota},$ then we have  
$$\prod_{B \in \ker{\boldsymbol \iota}} \mu_{\mathcal{NC}(\#(B))} (\alpha_B, \beta_B)=1$$
since $ \mu_{\mathcal{NC}(\#(B))} (\alpha_B, \beta_B)=1$ for every $B.$

Now suppose that $\alpha < \beta,$ then there should exist a $s\in [n]$ such that such that $\alpha_s$ is a strict refinement of $\beta_s$, i.e., $\alpha_s < \beta_s.$
Denote $S:= \{B \in \ker{\boldsymbol \iota: B \subseteq J_s}\}$ we have
\begin{equation*}
\begin{split}
\sum_{\substack{\eta \in \mathcal{NC}^{(\epsilon, {\boldsymbol \iota})}(k),\\ \alpha \leq \eta \leq \beta}} \prod_{B \in \ker{\boldsymbol \iota}}  \mu_{\mathcal{NC}(\#(B))} (\eta_B, \beta_B) 
&=  \sum_{\substack{\eta \in \mathcal{NC}^{(\epsilon, {\boldsymbol \iota})}(k),\\ \alpha \leq \eta \leq \beta}} \prod_{B \in S}  \mu_{\mathcal{NC}(\#(B))} (\eta_B, \beta_B) \cdot \prod_{B \notin S}  \mu_{\mathcal{NC}(\#(B))} (\eta_B, \beta_B)\\
& =  \sum_{\substack{\eta \in \mathcal{NC}^{(\epsilon, {\boldsymbol \iota})}(k),\\ \alpha \leq \eta \leq \beta}}  \mu_{\mathcal{NC}(k_s)} (\eta_s, \beta_s) \cdot \prod_{B \notin S}  \mu_{\mathcal{NC}(\#(B))} (\eta_B, \beta_B)\\
& = \sum_{ \alpha_s \leq \eta_s \leq \beta_s}  \mu_{\mathcal{NC}(k_s)} (\eta_s, \beta_s) \cdot \sum_{\substack{B \notin S,\\ \alpha_{B} \leq \eta_{B} \leq \beta_{B}}}\prod_{B \notin S}  \mu_{\mathcal{NC}(\#(B))} (\eta_B, \beta_B)\\
& =0,
\end{split}
\end{equation*}
where we used the fact that 
$$ \sum_{ \alpha_s \leq \eta_s \leq \beta_s}  \mu_{\mathcal{NC}(k_s)} (\eta_s, \beta_s)=0.$$
\end{proof}

\begin{rem}
For any $\alpha, \beta \in \mathcal{NC}^{(\epsilon, {\boldsymbol\iota})}(k)$, we have
\begin{equation}\label{eq:lattice}
[\alpha, \beta] \cong \prod_{B \in \ker{\iota}} [\alpha_B, \beta_B],
\end{equation}
where $[\alpha, \beta]: = \{\eta \in \mathcal{NC}^{(\epsilon, {\boldsymbol\iota})}(k): \alpha\leq \eta \leq \beta\}$ and $[\alpha_B, \beta_B]: =  \{\eta_B \in \mathcal{NC}(\#(B)): \alpha_B\leq \eta_B \leq \beta_B\}.$ Therefore, Proposition \ref{prop:mobius} is a natural corollary of \eqref{eq:lattice}. We refer to \cite{NS2006} for more details. 
\end{rem}


\subsection{Framework of noncommutative probability}

\noindent{\it Noncommutative probability space}--A noncommutative probability space \cite{NS2006} $(\mathcal{A},\phi)$ is an unital algebra $\mathcal{A}$ over $\mathbb{C}$, with a unital linear functional
\[\phi:\mathcal{A}\to\mathbb{C}; \;  \phi(\mathbf{1}_{\mathcal{A}})=1.\]
An element $a \in\mathcal{A}$ is called a noncommutative random variable, and it is called centered if $\phi(a)=0.$ Here are two examples of noncommutative probability spaces. 
\begin{enumerate}[{\rm (i)}]
\item $(L^\infty(\R, \mu), \mathbb{E})$, where  $\mu$ is a probability measure supporting on $\R,$ and  $\mathbb{E}$ is the expectation defined by $\mathbb{E} [f] := \int_{\R} f(t) d \mu(t)$.

\item $(\mathbb{M}_d(\mathbb{C}), {\rm tr})$, where $\mathbb{M}_d(\mathbb{C})$ is the algebra of $d \times d$ matrices, and ${\rm tr} := {\rm Tr}/d$ is the normalized trace.
\end{enumerate}

\vspace{2mm}

\noindent{\it $\epsilon$-Independence}--Independence plays a central role in probability theory. Roughly speaking, it provides a methodology to calculate the joint moments of random variables.  The notation of $\epsilon$-freely independence was firstly introduced by Mlotkowski \cite{M04} and later recovered by Speicher, and Wysoczanski \cite{SW16}. Here $\epsilon$ is a symmetric $m \times m$ matrix with entries to be $0$ or $1,$, and we follow the convention that the diagonal entries of $\epsilon$ are $0$'s. Fix a symmetric matrix $\epsilon,$ we denote $I_k^{(\epsilon)}$ by the set of all $k$-tuples of indices $(\iota_1, \ldots, \iota_k)$ from the integer set $[m]$ such that
whenever $\iota_i = \iota_j$ with $1 \leq i < j \leq k$ there is an $\ell$ with $i < \ell < j, \iota_i \neq \iota_\ell,$ and $\epsilon_{\iota_i, \iota_\ell}=0.$ 
Given a  noncommutative probability space $(\mathcal{A},\phi),$ let $\mathcal{A}_{1},\mathcal{A}_{2},\ldots,\mathcal{A}_{m}$ be a sequence of subalgebras of $\mathcal{A}$.  We call $\mathcal{A}_{1},\mathcal{A}_{2},\ldots,\mathcal{A}_{m}$ are $\epsilon$-freely independent, if the following holds:
\begin{enumerate}[{\rm (i)}]
\item $\mathcal{A}_i$ and $\mathcal{A}_j$ commute whenever $\epsilon_{i,j}=1$;
\item If for any centred elements $a_{j}\in \mathcal{A}_{\iota_{j}}$, $j=1,\ldots, k$, 
\begin{equation}\label{eq:free-independent}
\phi(a_{1}\cdots a_{k})=0
\end{equation}
whenever we have  $(\iota_1, \ldots, \iota_k) \in I_k^{(\epsilon)}.$ 
\end{enumerate}

A sequence of noncommutative random variables $a_1, \ldots, a_m$ are said to be $\epsilon$-freely independent if the subalgebras they generate are $\epsilon$-freely independent. Note that if the non-diagonal entries of $\epsilon$ are all $0$'s (resp. $1$'s),  then we reduce to Voiculescu's free independence (resp. classical independence).

\vspace{2mm}

\noindent{\it Moment-cumulant formulas for independent random variables}--Let $(\A,\varphi)$ be a noncommutative probability space. Given a family of random variables $a_{1},\ldots, a_{m}\in (\A,\phi)$, the mixed moments of the random variables $a_1, \ldots, a_m$ are given by  $\phi (a_{\iota_1} \cdots a_{\iota_k}),$ where ${\boldsymbol \iota} = (\iota_1, \ldots, \iota_k):[k] \rightarrow [m].$

Denote
\[\phi_{m}(a_{1},\ldots, a_{m}):=\phi(a_{1}\cdots a_{m}).\]
Then for any partition $\alpha \in \mathcal{P}(m)$, one can define \cite{NS2006}
\begin{equation}\label{eq:multi-moment}
\phi_{\alpha}[a_{1},\ldots,a_{m}]:=\prod_{V \in\alpha}\phi(V)[a_{1},\cdots,a_{m}],
\end{equation}
where $V$ is the block of $\alpha$ and 
\[\phi(V)[a_{1},\cdots,a_{m}]:=\phi_{\#(V)}(a_{i_{1}},\cdots,a_{i_{r}})\ \ \text{for}\ V=(i_{1},\cdots, i_{r}).\]
Hence $(\phi_{\alpha})_{\alpha \in \mathcal{NC}(m)}$ are multiplicative functionals on $\A^m$, namely, they factorize in a product according to the block structure of partitions \cite{NS2006}. 

Suppose that $\A_1, \ldots, \A_m$ are $\epsilon$-freely independent subalgebras of $\A$, then for arbitrary ${\boldsymbol \iota}: [k] \rightarrow [m]$, the mixed moment  can be represented as follows \cite{SW16, EPS18}:
\begin{equation}\label{eq:moment-cumulant}
\phi (a_1\cdots a_k) = \sum_{\alpha \in \mathcal{NC}^{(\epsilon, {\boldsymbol{\iota}})}(k)} \kappa_\alpha^{(\epsilon)}[a_1, \ldots, a_k],
\end{equation}
where $a_j \in \A_{\iota_j}, j=1, \ldots, k.$ The multilinear function $\kappa_\alpha^{(\epsilon)}[\cdots]$ is called $\epsilon$-free cumulant \cite{EPS18}. Note that if 
every non-diagonal entry of $\epsilon$ is $0$, Equation \eqref{eq:moment-cumulant} reduces to the following free cumulant-moment formula \cite{NS2006}
\begin{equation}\label{eq:free-cumulants}
 \phi (a_1 \cdots a_k) = \sum_{\substack{\alpha \in \mathcal{NC}(k), \\ \alpha \leq \ker{\boldsymbol \iota}}} \kappa_\alpha [a_1, \ldots, a_k],
\end{equation}
where $\{ \kappa_\alpha \}_{\alpha \in  \mathcal{NC}(k)}$ are called free cumulants. And if every non-diagonal entry of $\epsilon$ is $1$, we have the following classical cumulant-moment formula
\begin{equation}\label{eq:classical-cumulants}
 \phi (a_1 \cdots a_k) = \sum_{\substack{\alpha \in \mathcal{P}(k), \\ \alpha \leq \ker{\boldsymbol \iota}}}  k_\alpha [a_1, \ldots, a_k],
\end{equation}
where $\{ k_\alpha \}_{\alpha \in  \mathcal{P}(k)}$ are called classical cumulants. We note that $a_1, \ldots, a_m$ are freely independent in Equation \eqref{eq:free-cumulants} and classical independent in Equation \eqref{eq:classical-cumulants}, respectively. Indeed, the cumulants reflect the independence of random variables \cite{NS2006}, and the above moment-cumulant formulas indicate that the varieties of partitions yield different types of independence in the context of noncommutative probability space (see Table \ref{table:1}). 

\begin{table}[htb]
\begin{center}   
\caption{Partitions and Independence}  
\label{table:1} 
\begin{tabular}{|| m{3.8cm}<{\centering}| m{3.8cm}<{\centering}| m{3.8cm}<{\centering}||}   
\hline   \textbf{Independence} & \textbf{Partitions} & \textbf{Cumulants} \\   
\hline   Classical & All partitions & Classical cumulants \\ 
\hline   Free & Noncrossing partitions & Free cumulants  \\  
\hline   $\epsilon$-Free & $\epsilon$-Noncrossing partitions & $\epsilon$-Free cumulants  \\      
\hline   
\end{tabular}   
\end{center}   
\end{table}

Moreover, since we have made the convention that
the diagonal terms of $\epsilon$ are $0$'s, only free cumulants contributes in the sum \eqref{eq:moment-cumulant} (see \cite{SW16}[Theorem 5.2, Remark 5.4]), namely, for each $\alpha \in \mathcal{NC}^{(\epsilon, {\boldsymbol{\iota}})}(k)$
\begin{equation}
\begin{split}
\kappa_\alpha^{(\epsilon)} [a_1, \ldots, a_k]  : & = \kappa_\alpha [a_1, \ldots, a_k]\\
& = \prod_{V \in \alpha} \kappa_{\# (V)} [a_1, \ldots, a_k],
\end{split}
\end{equation}
where  $\kappa_{\alpha} [a_1, \ldots, a_k]$ is the product of the free cumulants for each block. 

Finally, it is known \cite{EPS18} that $\mathcal{NC}^{(\epsilon, {\boldsymbol \iota})}(k)$ is a lattice, and the $\epsilon$-free cumulants are multiplicative.
Therefore, due to the standard theory of convolution, one can obtain the following so-called M\"{o}bius inversion of Equation \eqref{eq:moment-cumulant}
\begin{equation}\label{eq:inversion-moment-cumulant}
\kappa_\alpha^{(\epsilon)} [a_1, \ldots, a_k]=\sum_{\substack{\beta \in \mathcal{NC}^{(\epsilon, {\boldsymbol{\iota}})}(k), \\ \beta \leq \alpha}}\phi_\beta [a_1, \ldots, a_k] \cdot \mu_{\mathcal{NC}^{(\epsilon, {\boldsymbol{\iota}})}(k)} (\beta, \alpha)
\end{equation}
for every $\alpha \in \mathcal{NC}^{(\epsilon, {\boldsymbol{\iota}})}(k).$


\section{Empirical distribution of the eigenvalues}

\noindent{\it Empirical distribution of the eigenvlues}--Let $\lambda_i, i=1, \ldots, d^n$ denote the eigenvalues of $H,$ and define \cite{AGZ09} the empirical distribution of the eigenvalues as the probability measure on $\mathbb{R}$ by 
\begin{equation}
\mu_H = \frac{1}{d^n} \sum_{i=1}^{d^n} \delta_{\lambda_i}.
\end{equation}

Let $(\A, \phi)$ be a noncommutative probability space. A random variable $a \in \A$ has a probability distribution $\mu$ on $\mathbb{R}$ if the following condition holds
\begin{equation}
\phi(a^k) = \int_{\mathbb{R}}  t^k \; d\mu(t), \;\; \text{for all} \; k\in\mathbb{N}.
\end{equation}
If we consider $H$ in the framework of noncommutative probability space $(\mathbb{M}_{d^n}(\mathbb{C}), {\rm tr})$, then we say $\mu_H$ converges weakly, almost surely, to a probability measure $\mu$ on $\mathbb{R}$ if almost surely, 
\begin{equation}
\lim_{d \rightarrow \infty} {\tr H^k}  = \int_{\mathbb{R}}  t^k \; d\mu(t), \;\; \text{for all} \; k\in\mathbb{N}.
\end{equation}

Define the semicircle distribution $\mu_{sc}$ as the probability distribution $\sigma(t) dt$ on $\mathbb{R}$ with density 
$$\sigma(t) = \frac{1}{2\pi} \sqrt{4-t^2} \cdot 1_{|t| \leq 2}.$$
Suppose that $H$ is globally sampled by Gaussian unitary ensemble (GUE); it is well known that $\mu_H$ converges weakly, almost surely, to $\mu_{sc}$ as the dimension $d$ goes to large \cite{Wig55}.

\vspace{2mm}

\noindent{\it $\epsilon$-free convolution}--convolution is an operation of probability measures. Suppose $a_1, \ldots, a_m \in \A$ associate with probability distribution $\mu_1, \ldots, \mu_m$, respectively. If $a_1, \ldots, a_m$ are classical (resp. freely) independent, then the classical (resp. freely) convolution of $\mu_1, \ldots, \mu_m$, denoted by $\mu_1\ast \cdots \ast \mu_m$ (resp. $\mu_1 \sqsum \cdots \sqsum \mu_m$), can be defined as follows \cite{NS2006}:
\begin{equation*}
\phi \left( \sum_{\iota=1}^m a_\iota \right)^k= \int_\mathbb{R} t^k \;  d (\mu_1 \ast \cdots \ast \mu_m) (t) \;  (\text{resp.} = \int_\mathbb{R} t^k \;  d (\mu_1 \sqsum \cdots \sqsum \mu_m) (t))
\end{equation*}
for all $k \in \mathbb{N}.$ Motivated by the above definition, we have the following definition

\begin{defn}\label{def:convolution}
Let $(\mathcal{A}, \phi)$ be a noncommutative probability space, let $a_\iota \in \A$ with probability distribution $\mu_\iota, \iota=1, \ldots, m.$ Suppose that $a_1, \ldots, a_m$ are $\epsilon$-freely independent for given 
$\epsilon.$ We define the $\epsilon$-free convolution of $\mu_1, \ldots, \mu_m$, denoted by $\mu_1 \sqsum_\epsilon \cdots \sqsum_\epsilon \mu_m,$ be the jointly distribution of $a_1+ \cdots + a_m,$ i.e.,
\begin{equation}
\begin{split}
\phi \left(\sum_{\iota=1}^m a_\iota \right)^k & = \int_{\mathbb{R}} t^k \; d\mu_{a_1+\cdots+a_m} (t)\\
& := \int_{\mathbb{R}} t^k \; d\left(\mu_1 \sqsum_\epsilon \cdots \sqsum_\epsilon \mu_m\right) (t)
\end{split}
\end{equation}
for all $k \in \mathbb{N}.$
\end{defn}
It is clear that if every non-diagonal entry of $\epsilon$ is $1$ (resp. every entry of $\epsilon$ is $0$)
then the $\epsilon$-free convolution reduces to the classical (resp. free) convolution. 

\vspace{2mm}

\noindent{\it Main results}--Case 1: Suppose that the local terms $H_{\iota}, \iota=1, \ldots, m$ are independently sampled by GUE. Namely, for every $\iota =1, \ldots, m$ we suppose
\begin{equation}\label{eq:Haar-H}
H_\iota =  G_\iota \otimes  \bigotimes_{s \notin \mathcal{K}_\iota} (\mathbb{C} \un_d),
\end{equation}
where $\{ G_\iota, \iota=1,\ldots, m\}$ is a family of independent GUEs with $G_\iota$ in $\bigotimes_{s\in \mathcal{K}_\iota} \mathbb{M}_d (\mathbb{C}).$ 
Note that $\mu_{H_\iota}$ converges weakly, almost surely, to $\mu_{sc}$ for each $\iota=1, \ldots, m.$

\begin{thm}\label{thm:GUE}
As $d \rightarrow \infty,$ $\mu_H$ converges weakly, almost surely, to $\sqsum^{(m)}_{\epsilon}\mu_{sc}$, if $H_{\iota}$'s are independently sampled by GUE. 
\end{thm}

\noindent Case 2: Suppose that $H_{\iota}$'s are independently sampled via Haar unitary invariant ensembles. Namely, for every $\iota =1, \ldots, m$ we suppose
\begin{equation}\label{eq:Haar-H}
H_\iota =   U_\iota A_\iota U_\iota^* \otimes \bigotimes_{s \notin \mathcal{K}_\iota} (\mathbb{C} \un_d),
\end{equation}
where $\{ U_\iota, \iota=1,\ldots, m\}$ is a family of independent unitaries with $U_\iota$ Haar-distributed in $\mathcal{U}\left( \bigotimes_{s\in \mathcal{K}_\iota} \mathbb{M}_d (\mathbb{C})\right)$ and 
$\{A_\iota, \iota=1, \ldots, m\}$ a family of deterministic Hermitian matrices with $A_\iota \in   \bigotimes_{s\in \mathcal{K}_\iota} \mathbb{M}_d (\mathbb{C}).$ 

\begin{assumption}\label{ass:1}
There are a sequence of compactly supported measures $\mu_1, \ldots, \mu_m$ on $\mathbb{R}$ such that $\mu_{A_\iota}$ weakly converges to $\mu_\iota$ for each $\iota=1, \ldots, m.$ Hence $\mu_{H_\iota}$ converges weakly, almost surely, to $\mu_\iota$ for each $\iota.$
\end{assumption}

\begin{thm}\label{thm:U}
As $d \rightarrow \infty,$  $\mu_H$ converges weakly, almost surely, to $\mu_1 \sqsum_{\epsilon} \cdots \sqsum_{\epsilon} \mu_m$, if $H_{\iota}$'s are independently sampled by the Haar unitary invariant ensemble which obeys Assumption \ref{ass:1}. 
\end{thm}

We sketch the proofs as follows and refer to Subsections \ref{subsec:moment}, \ref{subsec:almost-surely} and \ref{subsec:main} for the details. Firstly, we calculate the expectation of $k$th-moment ${\rm tr} (H^k)$ by using Wick's formula and the Weingarten formula for Cases 1 and 2, respectively. Indeed, it had been done independently in \cite{ML19}, and \cite{CC2015} for Case 2, and we only need to adapt their proofs in our setting. Secondly, we show that the leading order of the variance of ${\rm tr}(H^k)$ is $O(1/d^2).$ As a result, the Borel-Cantelli lemma can be applied, resulting in almost sure convergence. Finally, using the isomorphism between the $\epsilon$-noncrossing partition and permutation,  $\mu_H$ can be approximately represented by the $\epsilon$-free convolution of the large $d$ limit of $\mu_{H_\iota}$'s.

\subsection{Mixed moments of $H_1,\ldots, H_m$}\label{subsec:moment}

In this subsection, we calculate the mixed moments of $H_1, \ldots, H_m.$ The proofs follow  \cite{CC2015}, where the authors study the mixed moments of some matrix models that are asymptotically $\epsilon$-free. However, to be self-contained, we adapt their proofs to our settings. Before we start, we recall Wick's formula and Weingarten's formula, which enable us to calculate the 
mixed moments of GUE and Haar unitary ensemble, respectively. 

\begin{defn}\cite{NS2006, MS17}.
Let $G$ be a $d \times d$ matrix with entries $g_{i,j}$ where $\sqrt{d} g_{i,j}$ is a standard complex Gaussian random variable, i.e., $\mathbb{E}(g_{i,j})=0, \mathbb{E}\left(|g_{i,j}|^2 \right)=1/d$ and
\begin{enumerate}[{\rm (i)}]
\item $g_{i,j} = \overline{g_{i,j}};$
\item $\{{\rm Re}(g_{i,j})\}_{i\geq j} \cup \{{\rm Im}(g_{i,j})\}_{i> j}$ are independent.
\end{enumerate}
Then $G$ is called a GUE random matrix. 
\end{defn}

\begin{prop}\label{prop:Wick}\cite{NS2006, MS17} (Wick's formula).
Let $G=(g_{i,j})_{i,j=1}^d$ be a $d\times d$ GUE random matrix, then the expectation of the product of its entries can be calculated by the following Wick's formula:
\begin{equation}
\mathbb{E} \left( g_{i_1, j_1} \cdots g_{i_k j_k}\right)= \begin{cases} \sum_{\alpha \in \mathcal{P}_2(k)} \prod_{p \in [k]} \delta_{i_p, j_{\alpha(p)}} & \text{for}\; k \; \text{is even};\\
0 & \text{for}\; k \; \text{is odd}.
\end{cases}
\end{equation}
\end{prop}

\begin{prop}\label{prop:Weingarten}\cite{W78, C03} (Weingarten's formula).
Let $U=(u_{i,j})_{i,j=1}^d$ be a $d\times d$ unitary random matrix distributed according to the Haar measure of the group of $d\times d$ matrices $\mathcal{U}(d)$, then the expectation of the product of its entries can be calculated by the following Weingarten's formula:
\begin{equation}
\mathbb{E} \left( u_{i_1, j_1} \cdots u_{i_k j_k}\bar{u}_{i'_1,j'_1} \cdots \bar{u}_{i'_k, j'_k}\right)= \sum_{\sigma, \tau \in S_k} {\rm Wg} (\sigma \tau^{-1}, d) \cdot \prod_{p \in [k]}\delta_{i_p, i'_{\sigma(p)}}\delta_{j_p, j'_{\tau(p)}},
\end{equation}
where ${\rm Wg} (\sigma, d): S_k \rightarrow \mathbb{C}$ is called a Weingarten function.
\end{prop}
The following asymptotics of the Weingarten function is useful \cite{CS06}
\begin{equation}
{\rm Wg} (\sigma, d)= \mu (\sigma) d^{- (k+ |\sigma|)}\left(1+ O\left(d^{-2}\right) \right),
\end{equation}
where $\mu (\sigma)$ is a well defined function on $S_k$ and it can be shown \cite{NS2006} that 
$$\mu(\sigma \tau^{-1}) = \mu_{\mathcal{NC}}(\alpha, \beta),$$
where $\sigma= P_\alpha$ and $\tau = P_\beta.$

\begin{notation}
For given ${\boldsymbol{\iota}}: [k] \rightarrow [m]$ and $\sigma \in S^{({\boldsymbol \iota})}_k,$ 
\begin{enumerate}[{\rm (i)}]
\item We denote $\gamma: = \gamma_k = (1, \ldots, k).$
\item  We denote $\iota_B: = \iota_\ell$ for all $\ell \in B,$ and denote ${\rm \tilde{Wg}}(\sigma) := \prod_{B \in \ker{\boldsymbol \iota}} {\rm Wg}\left(\sigma|_B, d^{\#(\mathcal{K}_{\iota_B})}\right).$ 
\item ${\rm tr}_\sigma \left[ A_1, \ldots, A_k \right]: = \prod_{c \in \sigma} {\rm tr} \left( \prod_{i \in c} A_i\right),$ where $c$ is the cycle of $\sigma.$
\end{enumerate}
\end{notation}

\begin{prop}\label{prop:moment-GUE}
For the symmetric matrix $\epsilon$ given by \eqref{eq:epsilon}, if $H_\iota$'s are independently sampled GUEs, then the mixed moments of $H_1, \ldots, H_m$ 
is given as follows:
\begin{equation}\label{eq:moment-GUE}
\mathbb{E} \cdot {\rm tr} \left(H_{\iota_1} \cdots H_{\iota_k}\right) = \sum_{\alpha \in  \mathcal{P}_2^{({\boldsymbol{\iota}})}(k)}\prod_{s=1}^n d^{\# ( \gamma \cdot \alpha_s)-\frac{1}{2} k_s -1},
\end{equation}
for each ${\boldsymbol \iota}=(\iota_1, \ldots, \iota_k): [k]\rightarrow [m].$ 
\end{prop}

\begin{proof}
Note that the entries of $H_\iota= (H^{(\iota)}_{i,j})$ is given by 
$$ H^{(\iota)}_{i,j} := g^{(\iota)}_{i,j} \cdot \prod_{s \notin \mathcal{K}_\iota} \delta_{i[s], j[s]},$$
where $i, j$ are $n$-tuples described as $i = (i[1], \ldots, i[n]),$ and $g^{(\iota)}_{i,j}$ are the coefficients of $G_\iota.$
 We first suppose for convenience that $\mathbb{E}\left( \left|g_{i, j}^{(\iota)}\right|^2 \right)=1.$ Now, for any $k\geq 1$ we have
\begin{equation*}
\begin{split} 
 \mathbb{E} \cdot {\rm tr} (H_{\iota_1} H_{\iota_2} \cdots H_{\iota_k}) & = d^{-n}  \mathbb{E}  \cdot {\rm Tr} (H_{\iota_1} H_{\iota_2} \cdots H_{\iota_k})\\
& = d^{-n} \sum_{i_1, \ldots, i_k} \mathbb{E}  \left( H^{(\iota_1)}_{i_1, i_2} H^{(\iota_2)}_{i_2, i_3} \cdots H^{(\iota_k)}_{i_k i_1}\right)\\
& = d^{-n} \sum_{i_1, \ldots, i_k} \prod_{B \in \ker{\boldsymbol \iota}} \mathbb{E} \left( \prod_{p \in B}  H^{(\iota_p)}_{i_p, i_{\gamma(p)}}\right),
\end{split}
\end{equation*}
where $\gamma=(1,2,\ldots, k)$ is the full cycle of $[k].$
Note that $\iota_{p} = \iota_q$ for $p, q\in B,$ we denote $\iota_p = \iota_B$ for $p \in B.$ Thus it follows that
\begin{equation*}
\mathbb{E} \cdot {\rm tr} (H_{\iota_1} H_{\iota_2} \cdots H_{\iota_k})  = d^{-n} \sum_{i_1, \ldots, i_k} \prod_{B \in \ker{\boldsymbol \iota}} \mathbb{E} \left( \prod_{p \in B} g^{(\iota_B)}_{i_p, i_{\gamma(p)}}\right) \cdot  \prod_{p \in B}  \prod_{s \notin \mathcal{K}_{\iota_B}} \delta_{i_p [s], i_{\gamma(p)} [s]}. 
 \end{equation*}
 
By Wick's formula, $\mathbb{E} \left(  \prod_{p \in B} g_{i_p, i_{\gamma(p)}}^{(\iota_B)}\right)=0$ whenever $\# (B)$ is odd, and otherwise
\begin{equation*}
\begin{split}
\mathbb{E} \left(  \prod_{p \in B} g_{i_p, i_{\gamma (p)}}^{(\iota_B)}\right)
& = \sum_{\alpha \in \mathcal{P}_2(\# (B))} \prod_{p \in B} \prod_{s \in \mathcal{K}_{\iota_p}} \delta_{i_p[s], i_{\gamma \cdot \alpha (p)}[s]}.
\end{split}
\end{equation*} 
Since we need $\#(B)$ is even for every block $B \in \ker{\boldsymbol \iota} \in \mathcal{P}(k)$, therefore $\mathbb{E} \cdot {\rm tr} (\cdots) =0$ for $k$ is odd. 
Now for even $k,$ we are choosing independently pairings on each block $B \in \ker{\boldsymbol \iota}$ we may sum over all $\alpha \in \mathcal{P}_2^{\boldsymbol \iota}(k)$ outside of the product. Thus for even $k$, we have 
\begin{equation*}
\begin{split}
\mathbb{E} \cdot {\rm tr} (H_{\iota_1} H_{\iota_2} \cdots H_{\iota_k})   & = d^{-n}  \sum_{\alpha \in  \mathcal{P}_2^{({\boldsymbol{\iota}})}(k)} \sum_{i_1, \ldots, i_k} \prod_{B  \in \ker{\boldsymbol \iota}} \prod_{p \in B} \left( \prod_{s \in \mathcal{K}_{\iota_p}} \delta_{i_p[s], i_{\gamma \cdot \alpha (p)}[s]}\right)\cdot \left(\prod_{s \notin\mathcal{K}_{\iota_p}} \delta_{i_p[s], i_{\gamma(p)}[s]} \right)\\
& = d^{-n} \sum_{\alpha \in  \mathcal{P}_2^{({\boldsymbol{\iota}})}(k)} \sum_{i_1, \ldots, i_k}\prod_{\ell =1}^k \left( \prod_{s \in \mathcal{K}_{\iota_\ell}} \delta_{i_{\ell}[s], i_{ \gamma \cdot \alpha (\ell)}[s]}\right)\cdot \left(\prod_{s \notin\mathcal{K}_{\iota_\ell}} \delta_{i_\ell[s], i_{\gamma(\ell)}[s]} \right).
\end{split}
\end{equation*}
 
Let us consider the sum $\sum_{i_1, \ldots, i_k}\prod_{\ell =1}^k \left( \prod_{s \in \mathcal{K}_{\iota_\ell}} \delta_{i_{\ell}[s], i_{\gamma \cdot \alpha (\ell)}[s]}\right)\cdot \left(\prod_{s \notin\mathcal{K}_{\iota_\ell}} \delta_{i_\ell[s], i_{\gamma(\ell)}[s]} \right).$ In fact, these Dirac functions give the conditions for the indices that contribute to the sum. Similar to the proof in \cite{CC2015}, we will look at the conditions for each $s \in [n].$ We need $i_\ell [s] = i_{\gamma \cdot \alpha (\ell)}[s]$ when $s \in \mathcal{K}_{\iota_\ell},$ while $i_\ell [s] = i_{ \gamma(\ell)}[s]$ when $s \notin \mathcal{K}_{\iota_\ell}.$ That is, $i[s] = (\gamma \cdot \alpha_s) \cdot i[s],$ where we note that the action of permutation $\pi \in S_k$ on the indices $i[s] = (i_1[s], \ldots, i_k[s])$ is defined as $\pi \cdot i[s]: = (i_{\pi(1)}[s], \ldots, i_{\pi (k)}[s]).$ So we have
\begin{equation}\label{eq:sum}
\sum_{i_1, \ldots, i_k}\prod_{\ell =1}^k \left( \prod_{s \in \mathcal{K}_{\iota_\ell}} \delta_{i_{\ell}[s], i_{\alpha \cdot \gamma (\ell)}[s]}\right)\cdot \left(\prod_{s \notin\mathcal{K}_{\iota_\ell}} \delta_{i_\ell[s], i_{\gamma(\ell)}[s]} \right) = \prod_{s=1}^n d^{\# ( \gamma \cdot \alpha_s)}.
\end{equation}
To summarize, we have 
\begin{equation*}
\begin{split}
\mathbb{E} \cdot {\rm tr} (H_{\iota_1} H_{\iota_2} \cdots H_{\iota_k})  & = d^{-n} \sum_{\alpha \in  \mathcal{P}_2^{({\boldsymbol{\iota}})}(k)}\prod_{s=1}^n d^{\# (\gamma \cdot \alpha_s)} \\
&=  \sum_{\alpha \in  \mathcal{P}_2^{({\boldsymbol{\iota}})}(k)} \prod_{s=1}^n d^{\# (\gamma \cdot \alpha_s)-1}.
\end{split}
\end{equation*}

Now we return to the normalized entries such that $\mathbb{E}\left( \left|g_{i, j}^{(\iota)}\right|^2 \right)=d^{-\frac{1}{2}\# (\mathcal{K}_\iota)}.$
For normalized $H_\iota$ and even $k$, we have
\begin{equation}\label{eq:moment-H}
\begin{split}
\mathbb{E} \cdot {\rm tr} (H_{\iota_1} H_{\iota_2} \cdots H_{\iota_k}) & = \sum_{\alpha \in  \mathcal{P}_2^{({\boldsymbol{\iota}})}(k)} \prod_{\ell=1}^k d^{-\frac{1}{2} \#(\mathcal{K}_{\iota_\ell})}\cdot \prod_{s=1}^n d^{\# (\gamma \cdot \alpha_s)-1}\\
& = \sum_{\alpha \in  \mathcal{P}_2^{({\boldsymbol{\iota}})}(k)}\prod_{s=1}^n d^{\# (\gamma \cdot \alpha_s)-\frac{1}{2} k_s -1},
\end{split}
\end{equation}
where we have used the fact that 
$$\sum_{\ell=1}^k \# (\mathcal{K}_{\iota_\ell}) = \sum_{B \in \ker{\boldsymbol\iota}} \sum_{p \in B} \# (\mathcal{K}_{\iota_p}) = \sum_{s=1}^n k_s.$$
\end{proof}

The following proposition is one of the main results in \cite{CC2015}. We keep the proof for the convenience of the readers after adapting the notations and adding a few details.
\begin{prop}\label{prop:moment-U}\cite{CC2015}[Theorem 6].
For the symmetric matrix $\epsilon$ given by \eqref{eq:epsilon}, if $H_{\iota}$'s are independently sampled by a Haar unitary invariant ensemble, then the mixed moments of $H_1, \ldots, H_m$ 
are given as follows:
\begin{equation}
\mathbb{E} \cdot {\rm tr} \left(H_{\iota_1} \cdots H_{\iota_k}\right) =\sum_{\sigma, \tau \in S^{({\boldsymbol{\iota}})}_k} {\rm \tilde{Wg}}(\sigma \tau^{-1}) \cdot {\rm tr}_{\tau}\left[ A_{\iota_1}, \ldots, A_{\iota_k}\right]\cdot \prod_{s=1}^n d^{ \#_{k_s} (\tau_s)+ \# (\sigma_s^{-1}\gamma)-1},
\end{equation}
for each ${\boldsymbol \iota}=(\iota_1, \ldots, \iota_k): [k]\rightarrow [m].$
\end{prop}

\begin{proof}
Note that the entries of $H_\iota= (H^{(\iota)}_{i,j})$ are given by 
$$ H^{(\iota)}_{i,j} := \sum_{p,q} a_{p, q}^{(\iota)} u^{(\iota)}_{i, p}\bar{u}_{j,q}^{(\iota)},$$ 
where $i, j, p, q$ are $n$-tuples described as $i = (i[1], \ldots, i[n]),$ and $a_{p, q}^{(\iota)}$ and $u^{(\iota)}_{p, q}$ are the coefficients of $A_\iota$ and $U_\iota,$ respectively.
For any $k\geq 1$, we have
\begin{equation*}
\begin{split} 
\mathbb{E} \cdot {\rm tr} \left(H_{\iota_1} \cdots H_{\iota_k}\right) 
& = d^{-n} \sum_{i_1, \ldots, i_k} \mathbb{E}  \left( H^{(\iota_1)}_{i_1, i_2} H^{(\iota_2)}_{i_2, i_3} \cdots H^{(\iota_k)}_{i_k i_1}\right)\\
& = d^{-n} \sum_{i_1, \ldots, i_k} \prod_{B \in \ker{\boldsymbol \iota}} \mathbb{E} \left( \prod_{t \in B}  H^{(\iota_t)}_{i_t, i_{\gamma(t)}}\right).
\end{split}
 \end{equation*}

We  use the Weingarten formula to calculate the term $\mathbb{E} \left( \prod_{t \in B}  H^{(\iota_t)}_{i_t, i_{\gamma(t)}}\right)$.  We denote $\iota_t= \iota_B$ for all $t \in B.$
\begin{equation*}
\begin{split}
\mathbb{E} \left( \prod_{t \in B}  H^{(\iota_t)}_{i_t, i_{\gamma(t)}}\right) & = \mathbb{E} \left( \prod_{t \in B}  \sum_{p_t,q_t} a_{p_t, q_t}^{(\iota_t)} u^{(\iota_t)}_{i_t, p_t}\bar{u}_{i_{\gamma(t)},q_t}^{(\iota_t)}\right)\\
& = \sum_{p_t, q_t, t \in B} \mathbb{E} \left( \prod_{t \in B}  a_{p_t, q_t}^{(\iota_t)} u^{(\iota_t)}_{i_t, p_t}\bar{u}_{i_{\gamma(t)},q_t}^{(\iota_t)}\right)\\
& =  \sum_{p_t, q_t, t \in B}\left( \prod_{t \in B} a_{p_t, q_t}^{(\iota_t)}\right) \cdot 
 \Big( \sum_{\sigma, \tau\in S_B} {\rm Wg}\left(\sigma\tau^{-1}, d^{\#(\mathcal{K}_{\iota_B})}\right) \times \\ 
&\times \prod_{t \in B} \prod_{s \in \mathcal{K}_{\iota_t}} \delta_{i_t[s], i_{\sigma^{-1}\gamma(t)}[s]} \delta_{p_t[s], q_{\tau^{-1}(t)}[s]}\cdot \prod_{s \notin \mathcal{K}_{\iota_t}} \delta_{i_t[s], i_{\gamma(t)}[s]} \Big)\\
& =\sum_{\sigma, \tau\in S_B}{\rm Wg}\left(\sigma\tau^{-1}, d^{\#(\mathcal{K}_{\iota_B})}\right) \sum_{p_t, t \in B}\prod_{t \in B} \prod_{s \in \mathcal{K}_{\iota_t}} a^{(\iota_t)}_{p_t[s], p_{\tau(t)}[s]}\delta_{i_t [s], i_{\sigma^{-1}\gamma(t)}[s]}\cdot \prod_{s \notin \mathcal{K}_{\iota_t}} \delta_{i_t[s], i_{\gamma(t)}[s]}. 
\end{split}
\end{equation*}
Since we are choosing permutations on each $B$ independently, we may instead sum over all $\sigma, \tau \in S^{({\boldsymbol{\iota}})}_k$ outside of the product. Therefore it follows that
\begin{equation*}
\begin{split}
\mathbb{E} \cdot {\rm tr} \left(H_{\iota_1} \cdots H_{\iota_k}\right)  & = d^{-n} \sum_{\sigma, \tau \in  S^{({\boldsymbol{\iota}})}_k}{\rm \tilde{Wg}}(\sigma \tau^{-1})
 \cdot \sum_{\substack{i_1, \ldots, i_k;\\ p_1,\ldots, p_k}}
\prod_{B \in \ker{\boldsymbol \iota}} \prod_{t \in B}\prod_{s \in \mathcal{K}_{\iota_t}} a^{(\iota_t)}_{p_t[s], p_{\tau(t)}[s]}\delta_{i_t[s], i_{\sigma^{-1}\gamma(t)}[s]}\cdot \prod_{s \notin\mathcal{K}_{\iota_t}} \delta_{i_t[s], i_{\gamma(t)}[s]} \\
& = d^{-n} \sum_{\sigma, \tau \in  S^{({\boldsymbol{\iota}})}_k}{\rm \tilde{Wg}}(\sigma \tau^{-1}) \cdot \sum_{\substack{i_1, \ldots, i_k;\\ p_1,\ldots, p_k}}
\prod_{\ell=1}^k \prod_{s \in \mathcal{K}_{\iota_\ell}} a^{(\iota_\ell)}_{p_\ell[s], p_{\tau(\ell)}[s]}\delta_{i_\ell[s], i_{\sigma^{-1}\gamma(\ell)}[s]}\cdot \prod_{s \notin \mathcal{K}_{\iota_\ell}} \delta_{i_\ell[s], i_{\gamma(\ell)}[s]}.
\end{split}
\end{equation*}

Again, let us consider the conditions for the indices that contributes in the sum $\sum_{\substack{i_1, \ldots, i_k;\\ p_1,\ldots, p_k}} [\cdots].$
Firstly, similar to Equation \eqref{eq:sum}, we have 
$$\sum_{i_1, \ldots, i_k}
\prod_{\ell=1}^k \prod_{s \in \mathcal{K}_{\iota_\ell}} \delta_{i_\ell[s], i_{\sigma^{-1}\gamma(\ell)}[s]}\cdot \prod_{s \notin \mathcal{K}_{\iota_\ell}} \delta_{i_\ell[s], i_{\gamma(\ell)}[s]}=\prod_{s=1}^n d^{ \# (\sigma_s^{-1}\gamma)}.$$
Next, we turn to consider the sum  $\sum_{p_1,\ldots, p_k} \prod_{\ell=1}^k \prod_{s \in \mathcal{K}_{\iota_\ell}} a^{(\iota_\ell)}_{p_\ell[s], p_{\tau(\ell)}[s]}.$ For each block $B \in \ker{\boldsymbol\iota}$, if we look at the sum based on $\tau|_B$, then we have a trace (over $\otimes_{s \in \mathcal{K}_\iota} \mathbb{M}_d(\mathbb{C})$) of $A_\iota$'s respect to $\tau|_B.$ Namely, 
\begin{equation*}
\begin{split}
\sum_{p_1,\ldots, p_k} \prod_{\ell=1}^k \prod_{s \in \mathcal{K}_{\iota_\ell}} a^{(\iota_\ell)}_{p_\ell[s], p_{\tau(\ell)}[s]} & = \prod_{B \in \ker{\boldsymbol \iota}} {\rm Tr}_{\tau|_B}[A_1, \ldots. A_k]\\
& = \prod_{B \in \ker{\boldsymbol \iota}} d^{\#_B (\tau|_B) \cdot \#(\mathcal{K}_{\iota_B})} {\rm tr}_{\tau|_B}[A_1, \ldots. A_k]\\
& = {\rm tr}_\tau [A_1, \ldots, A_k] \cdot \prod_{s=1}^n d^{\#_{k_s} (\tau_s)},
\end{split}
\end{equation*}
where we have used the fact that
$$\sum_{s=1}^n \#_{k_s} (\tau_s) = \sum_{B \in \ker {\boldsymbol \iota}} \#_B (\tau|_B) \cdot \#(\mathcal{K}_{\iota_B}),$$
since $\#(\mathcal{K}_{\iota_B})$ is the number of occurrences that the block $B$ appears in $J_s.$

To sum up, we have
\begin{equation}\label{eq:moment-H-U}
\begin{split}
\mathbb{E} \cdot {\rm tr} \left(H_{\iota_1} \cdots H_{\iota_k}\right) & = d^{-n}  \sum_{\sigma, \tau \in  S^{({\boldsymbol{\iota}})}_k}{\rm \tilde{Wg}}(\sigma \tau^{-1}) \cdot {\rm tr}_{\tau}\left[ A_{\iota_1}, \ldots, A_{\iota_k}\right]\cdot 
\prod_{s=1}^n d^{ \#_{k_s} (\tau_s)+ \# (\sigma_s^{-1}\gamma)}\\
& =  \sum_{\sigma, \tau \in S^{({\boldsymbol{\iota}})}_k} {\rm \tilde{Wg}}(\sigma \tau^{-1}) \cdot {\rm tr}_{\tau}\left[ A_{\iota_1}, \ldots, A_{\iota_k}\right]\cdot \prod_{s=1}^n d^{ \#_{k_s} (\tau_s)+ \# (\sigma_s^{-1}\gamma)-1}.
\end{split}
\end{equation}
\end{proof}

We end this subsection with the following remark. 
\begin{rem}
By Proposition \ref{prop:moment-U}, the local terms $H_1, \ldots, H_m$ that satisfies conditions in Case II are asymptotic $\epsilon$-free (see \cite{CC2015}[Theorem 7]). Moreover, because the GUE random matrix is Haar unitary invariant, Proposition \ref{prop:moment-U} could be used to induce the GUE model's asymptotic $\epsilon$-freeness. However, we would like to derive an explicit formula for the moments of the GUE model to access the almost sure convergence, which is one of the main results of our paper.
\end{rem}


\subsection{Almost sure convergence}\label{subsec:almost-surely}

\begin{notation}
For given $\tilde{\boldsymbol{\iota}}=(\iota_1, \ldots, \iota_k, \iota_{k+1}, \ldots, \iota_{2k}): [2k] \rightarrow m.$ 
\begin{enumerate}[{\rm (i)}]
\item Denote $\gamma_1= (1,2, \ldots, k)$, $\gamma_2 = (k+1, k+2, \ldots, 2k)$, and  $\delta = \gamma_1 \times \gamma_2 \in S_{2k}.$ 
\item For each string $s \in [n],$ we denote $J_{1, s}: = J_s \cap [1,k]$ and $J_{2, s} := J_s \cap [k+1, 2k].$ Moreover, $k_{i,s} : = \# J_{i, s}, \; i=1,2.$
\item Any $\sigma \in S_{2k}^{(\tilde{\boldsymbol{\iota}})}$ induces permutations $\sigma_{i,s} \in S_{J_{i,s}},$ $i=1,2.$ 
\item Denote $\alpha_{i,s}:= \alpha|_{J_{i,s}},  \; i=1.2.$ 
\end{enumerate}
\end{notation}

A partition $\alpha \in \mathcal{P} (2k)$ is called connected if there is a block $B \in \alpha$ such that $B \cap [1,k] \neq \emptyset$ and $B \cap [k+1,2k] \neq \emptyset.$ A partition $\alpha \in \mathcal{P}(2k)$ is connected
if and only if $P_\alpha \in S_{2k}$ is connected, i.e., $P_\alpha$ and $\delta$ generates a transitive subgroup in $S_{2k}.$

\begin{notation}
For given $\tilde{\boldsymbol{\iota}}: [2k] \rightarrow m,$ we denote $\mathcal{P}_{2,c}^{(\tilde{\boldsymbol{\iota}})}(2k)$ by the set of all connected pair partitions of $\mathcal{P}_2^{(\tilde{\boldsymbol{\iota}})}(2k)$, and by $S_{2k,c}^{(\tilde{\boldsymbol \iota})}$ the set of all connected permutations of $S_{2k}^{(\tilde{\boldsymbol \iota})}$.
\end{notation}

\begin{prop}\label{prop:almost-GUE}
For the symmetric matrix $\epsilon$ given by \eqref{eq:epsilon}, if $H_\iota$'s are independently sampled by GUE, then almost sure we have
\begin{equation}
\lim_{d \rightarrow \infty} {\rm tr} \left(H_{\iota_1} \cdots H_{\iota_k}\right) = \sum_{\alpha \in  \mathcal{NC}_2^{(\epsilon,{\boldsymbol{\iota}})}(k)}1
\end{equation}
for each ${\boldsymbol \iota}=(\iota_1, \ldots, \iota_k): [k]\rightarrow [m].$ 
\end{prop}

\begin{proof}
The proof follows the ideas in \cite{MSS2007}. Firstly, note that for any pair partition $\alpha_s,$ we always have $\# (\alpha_s \gamma)=\#_{k_s} (\alpha_s \gamma_s) \leq \frac{k_s}{2} +1$, the equality holds whenever $\alpha_s$ is non-crossing. Therefore for each ${\boldsymbol \iota}=(\iota_1, \ldots, \iota_k): [k]\rightarrow [m]$ we have
\begin{equation}\label{eq:moment-GUE}
 \mathbb{E} \cdot {\rm tr} \left(H_{\iota_1} \cdots H_{\iota_k}\right)= \sum_{\alpha \in  \mathcal{NC}_2^{(\epsilon,{\boldsymbol{\iota}})}(k)} 1 + O\left( \frac{1}{d}\right),
\end{equation}
where we have used Equation \eqref{eq:pair-trans}.
Hence
\begin{equation}
\lim_{d \rightarrow \infty} \mathbb{E} \cdot {\rm tr} \left(H_{\iota_1} \cdots H_{\iota_k}\right)= \sum_{\alpha \in  \mathcal{NC}_2^{(\epsilon,{\boldsymbol{\iota}})}(k)} 1,
\end{equation}

For the almost sure convergence, by using the Borel-Cantelli lemma, it suffices to show that
\begin{equation}\label{eq:Borel-Cantelli}
\sum_{d=1}^\infty \mathbb{E}\left( {\rm tr}\left( H_{\iota_1} \cdots H_{\iota_k} \right)- \mathbb{E} \cdot {\rm tr} \left(  H_{\iota_1} \cdots H_{\iota_k}  \right) \right)^2 < \infty.
\end{equation}
To this end, let us consider the following expectation of the product of normalized traces: 
\begin{equation*}
\begin{split}
\mathbb{E} \cdot {\rm tr}^2 \left( H_{\iota_1} \cdots H_{\iota_k}  \right) 
& = d^{-2n} \sum_{\substack{j_1, \ldots, j_{k},\\ j_{k+1}, \ldots, j_{2k}}}\mathbb{E}\left( H^{(\iota_1)}_{j_1,j_2}\cdots H^{(\iota_k)}_{j_k,j_1}\cdot H^{(\iota_{k+1})}_{j_{k+1}, j_{k+2}}\cdots H^{(\iota_{2k})}_{j_{2k},j_{k+1}} \right)\\
& = d^{-2n} \sum_{j_1,\ldots, j_{2k}}\mathbb{E}\left( \prod_{\ell=1}^{2k} H^{(\iota_\ell)}_{j_\ell,j_{\delta(\ell)}} \right).
\end{split}
\end{equation*}
Similar to the derivation of Equation \eqref{eq:moment-H}, we have
\begin{equation*} 
\begin{split}
\mathbb{E} \cdot {\rm tr}^2 \left( H_{\iota_1} \cdots H_{\iota_k}  \right)
& =\sum_{\alpha \in \mathcal{P}_2^{(\tilde{\boldsymbol{\iota}})}(2k)} \prod_{\ell=1}^{2k} d^{-\frac{1}{2}\mathcal{K}_{\iota_\ell}}\prod_{s=1}^n d^{\#(\delta \cdot \alpha_s)-2}\\
& = \sum_{\alpha \in \mathcal{P}_2^{(\tilde{\boldsymbol{\iota}})}(2k)}\prod_{s=1}^n d^{\#( \delta \cdot \alpha_s)-\frac{1}{2}k_s-2}.
\end{split}
\end{equation*}

Let $\boldsymbol{\iota}_1$ and $\boldsymbol{\iota}_2$ be the restrictions of $\tilde{\boldsymbol{\iota}}$ to $[1, k]$ and to $[k+1, 2k],$ respectively. It follows that
\begin{equation*}
\begin{split}
\mathbb{E} \cdot {\rm tr}^2 \left( H_{\iota_1} \cdots H_{\iota_k}  \right) - \left(\mathbb{E} \cdot {\rm tr} \left( H_{\iota_1} \cdots H_{\iota_k}\right) \right)^2 
& = \sum_{\alpha \in \mathcal{P}_2^{(\tilde{\boldsymbol{\iota}})}(2k)} \prod_{s=1}^n d^{\#(\delta \cdot \alpha_s)-\frac{1}{2}k_s-2}\\
& - \sum_{\substack{\alpha_1 \in \mathcal{P}_2^{({\boldsymbol\iota_1})}(k),\\ \alpha_2 \in \mathcal{P}_2^{({\boldsymbol\iota_2})}(k)}} \prod_{s=1}^n d^{\#(\gamma_1 \cdot \alpha_{1,s})
+ \#(\gamma_2 \cdot \alpha_{2,s})-\frac{1}{2}k_s-2}\\
& =  \sum_{\alpha \in \mathcal{P}_{2,c}^{(\tilde{\boldsymbol{\iota}})}(2k)}\prod_{s=1}^n d^{\#(\delta \cdot \alpha_s)-\frac{1}{2}k_s-2}\\
& =  \sum_{\alpha \in \mathcal{P}_{2,c}^{(\tilde{\boldsymbol{\iota}})}(2k)}\prod_{s \in S_1} d^{(\cdots)} \cdot \prod_{s \in S_2} d^{(\cdots)},
\end{split}
\end{equation*}
where we denote $S_1: = \{s\in [n]: \alpha_s \; \text{is connected}\}$ and $S_2:= \{s\in [n]: \alpha_s \; \text{is disconnected}\}.$ We note that $S_1  \neq \emptyset,$ otherwise $\alpha$ could not be
connected.

For any  $s \in S_1,$ by Formula \eqref{eq:Euler} we have
\begin{equation*}
\begin{split}
\#(\alpha_s \delta) &= \#_{k_s}(\alpha_s\delta_s)= k_s + 2(1-g) - \#_{k_s} (\alpha_s) -\#_{k_s} (\delta_s)\\
& = k_s+ 2(1-g) - \frac{1}{2}k_s- 2\\
& = \frac{1}{2}k_s -2g.
\end{split}
\end{equation*}
And for any $s\in S_2,$ we have $\alpha_s = \alpha_{1,s} \times \alpha_{2,s}$ and 
\begin{equation*}
\begin{split}
\#(\alpha_s  \delta)-\frac{1}{2}k_s-2  = \#(\alpha_{1,s} \gamma_1)+\#(\alpha_{2,s} \gamma_2) -\frac{1}{2}k_s-2 \leq 0.
\end{split}
\end{equation*}
The equality holds when $\pi_s$ is noncrossing.

In summary, the leading order of terms in the above sum is $O\left (d^{-2\#(S_1)} \right)$ (by letting $g=0$), thus finally we have 
$$\mathbb{E} \cdot {\rm tr}^2 \left( H_{\iota_1} \cdots H_{\iota_k}  \right) - \left(\mathbb{E} \cdot {\rm tr} \left( H_{\iota_1} \cdots H_{\iota_k}\right) \right)^2 = O(d^{-2}),$$
which implies \eqref{eq:Borel-Cantelli}.
\end{proof}

\begin{prop}\label{prop:almost-U}
For the symmetric matrix $\epsilon$ given by \eqref{eq:epsilon}, if $H_{\iota}$'s are independently sampled by a Haar unitary invariant ensemble which obeys Assumption \ref{ass:1}, then almost surely we have 
\begin{equation}
\begin{split}
\lim_{d \rightarrow \infty} {\rm tr} \left(H_{\iota_1} \cdots H_{\iota_k}\right) =  \sum_{\substack{\sigma, \tau \in S^{(\epsilon,{\boldsymbol{\iota}})}_{NC}(\gamma);\\ \tau \leq \sigma }} \mu(\sigma \tau^{-1}) \cdot \lim_{d\rightarrow \infty} {\rm tr}_{\tau}\left[ A_{\iota_1}, \ldots, A_{\iota_k}\right].
\end{split}
\end{equation}
\end{prop}

\begin{proof}
Recall that for $\sigma \in S^{({\boldsymbol{\iota}})}_k$ we have
\begin{equation*}
\begin{split}
{\rm \tilde{Wg}}(\sigma )
& =  \prod_{B \in \ker{\boldsymbol \iota}} \mu(\sigma|_B) d^{-\#(\mathcal{K}_{\iota_B})\cdot (\# (B)+ |\sigma|_B)}\left(1+ O\left(d^{-2\#(\mathcal{K}_{\iota_B})}\right) \right)\\
& =  \mu(\sigma) \prod_{s=1}^n d^{- (k_s+ |\sigma_s|)}(1+ O(d^{-2})),
\end{split}
\end{equation*}
where $\mu(\sigma):= \prod_{B \in \ker{\boldsymbol \iota}} \mu(\sigma_B).$

Combining Equation \eqref{eq:moment-H-U}, we obtain
\begin{equation}\label{eq:moment-H-U-2}
\begin{split}
\mathbb{E} \cdot {\rm tr} \left(H_{\iota_1} \cdots H_{\iota_k}\right) =   \sum_{\sigma, \tau \in S^{({\boldsymbol{\iota}})}_k} \mu(\sigma \tau^{-1}) \cdot {\rm tr}_{\tau}\left[ A_{\iota_1}, \ldots, A_{\iota_k}\right]\cdot 
\prod_{s=1}^n d^{ \#_{k_s} (\tau_s)+ \# (\sigma_s^{-1}\gamma)-k_s-|\sigma_s\tau_s^{-1}|-1}(1+ O(d^{-2})).
\end{split}
\end{equation}
Given $\sigma,\tau \in S^{({\boldsymbol{\iota}})}_k$ we have 
\begin{equation*}
\begin{split}
 \#_{k_s} (\tau_s)+ \# (\sigma_s^{-1}\gamma)-k_s-|\sigma_s\tau_s^{-1}|-1 & = k_s-1-(|\tau_s+ |\sigma_s\tau_s^{-1}| +|\sigma_s^{-1}\gamma_s|)\\
& \leq k_s-1-|\gamma_s|\leq 0.
\end{split}
\end{equation*}
The equality holds whenever $\sigma_s, \tau_s$ is on the geodesic path $1_s-\tau_s-\sigma_s-\gamma_s$ in $S_{k_s}.$ Thus Equation \eqref{eq:moment-H-U-2} induces that
\begin{equation}
\begin{split}
\lim_{d \rightarrow \infty} \mathbb{E} \cdot {\rm tr} \left(H_{\iota_1} \cdots H_{\iota_k}\right) & =  \sum_{\substack{\sigma, \tau \in S^{({\boldsymbol{\iota}})}_k;\\ 1_s-\tau_s-\sigma_s-\gamma_s \;\text{for all} \; s} } \mu(\sigma \tau^{-1}) \cdot \lim_{d\rightarrow \infty}{\rm tr}_{\tau}\left[ A_{\iota_1}, \ldots, A_{\iota_k}\right]\\
 & =  \sum_{\substack{\sigma, \tau \in S^{(\epsilon,{\boldsymbol{\iota}})}_{NC}(\gamma);\\ \tau \leq \sigma }} \mu(\sigma \tau^{-1}) \cdot \lim_{d\rightarrow \infty}{\rm tr}_{\tau}\left[ A_{\iota_1}, \ldots, A_{\iota_k}\right].
\end{split}
\end{equation}
Note that due to  Assumption \ref{ass:1}, the limit $\lim_{d\rightarrow \infty} {\rm tr}_\tau \left[ A_{\iota_1}, \ldots, A_{\iota_k}\right]$ always exists.

Now we turn to show the almost sure convergence. Similar to the proof in Proposition \ref{prop:almost-GUE}, we have
\begin{equation*}
\begin{split}
\mathbb{E} \cdot {\rm tr}^2 \left( H_{\iota_1} \cdots H_{\iota_k}  \right) &- \left(\mathbb{E} \cdot {\rm tr} \left( H_{\iota_1} \cdots H_{\iota_k}\right) \right)^2  \\
& = \sum_{\sigma, \tau \in S^{(\tilde{\boldsymbol{\iota}})}_{2k}} {\rm \tilde{Wg}}(\sigma \tau^{-1}) \cdot {\rm tr}_{\tau}\left[ A_{\iota_1}, \ldots, A_{\iota_{2k}}\right]\cdot \prod_{s=1}^n d^{ \#_{k_s} (\tau_s)+ \# (\sigma^{-1}_s \delta)-2}\\
&- \sum_{\substack{\sigma_1, \tau_1 \in S^{({\boldsymbol{\iota}_1})}_{k};\\
\sigma_2, \tau_2 \in S^{({\boldsymbol{\iota}_2})}_{k}}} {\rm \tilde{Wg}}(\sigma_1 \tau_1^{-1}){\rm \tilde{Wg}}(\sigma_2 \tau_2^{-1})  \times\\
& \times {\rm tr}_{\tau_1\times \tau_2}\left[ A_{\iota_1}, \ldots, A_{\iota_{2k}}\right]\cdot 
d^{ \#_{k_{1,s}} (\tau_1)+ \# (\sigma_1^{-1}\gamma_1)+  \#_{k_{2,s}} (\tau_2)+ \# (\sigma_2^{-1}\gamma_2)-2}\\
& = \sum_{\sigma, \tau \in S^{(\tilde{\boldsymbol{\iota}})}_{2k,c}} {\rm \tilde{Wg}}(\sigma \tau^{-1}) \cdot {\rm tr}_{\tau}\left[ A_{\iota_1}, \ldots, A_{\iota_{2k}}\right]\cdot \prod_{s=1}^n d^{ \#_{k_s} (\tau_s)+ \# (\sigma^{-1}\delta)-2}\\
& = \sum_{\sigma, \tau \in S^{(\tilde{\boldsymbol{\iota}})}_{2k,c}}\mu(\sigma \tau^{-1}) \cdot {\rm tr}_{\tau}\left[ A_{\iota_1}, \ldots, A_{\iota_{2k}}\right]\cdot \prod_{s=1}^n d^{ \#_{k_s} (\tau_s)+ \# (\sigma_s^{-1}\delta)-|\sigma_s\tau_s^{-1}|-k_s-2}\\
& = \sum_{\sigma, \tau \in S^{(\tilde{\boldsymbol{\iota}})}_{2k,c}}\mu(\sigma \tau^{-1}) \cdot {\rm tr}_{\tau}\left[ A_{\iota_1}, \ldots, A_{\iota_{2k}}\right]  \cdot\prod_{s\in S_1} d^{(\cdots)}\cdot \prod_{s\in S_2} d^{(\cdots)}\cdot \prod_{s\in S_3} d^{(\cdots)}\cdot \prod_{s\in S_4} d^{(\cdots)},
\end{split}
\end{equation*}
where we denote $S_1:=\{s\in [n]: \sigma_s, \tau_s \; \text{are connected}\}, S_2:=\{s\in [n]: \sigma_s \; \text{is connected and } \tau_s \;\text{is disconnected}\}, S_3:=\{s\in [n]: \tau_s \; \text{is connected and} \; \sigma_s \;\text{is disconnected}\},$ and $S_4:=\{s\in [n]: \sigma_s, \tau_s \; \text{are disconnected}\}.$

For any $s\in S_1$ we have
\begin{equation*}
\begin{split}
 \#_{k_s} (\tau_s)+ \# (\sigma_s^{-1}\delta)-|\sigma_s\tau_s^{-1}|-k_s-2& = k_s-2-(|\tau_s|+ |\sigma_s\tau_s^{-1}| +|\sigma^{-1}_s\delta_s|)\\
& \leq k_s-2-(|\sigma_s|+ |\sigma^{-1}_s \delta_s|)\\
& \leq k_s-2-k_s=-2,
\end{split}
\end{equation*}
where for the first inequality we have used the triangle inequality $|\sigma_s|\leq |\tau_s|+ |\sigma_s\tau_s^{-1}|,$ and for the second one we have used the fact that $k_s \leq |\sigma_s|+ |\sigma^{-1}_s\delta_s|$ since $\sigma_s$ is connected.

For any $s\in S_2$ we have
\begin{equation*}
\begin{split}
 \#_{k_s} (\tau_s)+ \# (\sigma_s^{-1}\delta)-|\sigma_s\tau_s^{-1}|-k_s-2& = k_s-2-(|\tau_s|+ |\sigma_s\tau_s^{-1}| +|\sigma^{-1}_s\delta_s|)\\
&< k_s-2-(|\sigma_s|+ |\sigma^{-1}_s \delta_s|)\\
& \leq k_s-2-k_s=-2.
\end{split}
\end{equation*}
We noted that for $s\in S_2,$ $|\sigma_s|< |\tau_s|+ |\sigma_s\tau_s^{-1}|.$ Since the connectedness of $\sigma_s$ and $\tau_s$ is different, they can not lies in a geodesic path. The case for $s\in S_3$ is similar.

For any $s\in S_4,$ we have $\sigma_s = \sigma_{1,s} \times \sigma_{2,s}, \tau_s = \tau_{1,s} \times \tau_{2,s}$ and 
\begin{equation*}
\begin{split}
 \#_{k_s} (\tau_s)+ \# (\sigma_s^{-1}\delta)-|\sigma_s\tau_s^{-1}|-k_s-2& = k_s-2-(|\tau_s|+ |\sigma_s\tau_s^{-1}| +|\sigma^{-1}_s\delta_s|)\\
&\leq k_s-2-(|\sigma_s|+ |\sigma^{-1}_s \delta_s|)\\
& = k_s-2-(|\sigma_{1,s}|+|\sigma_{2,s}| + |\sigma^{-1}_{1,s} \gamma_s|+|\sigma^{-1}_{2,s} \gamma_s|)\\
& \leq k_s-2-(k_s-2)=0.
\end{split}
\end{equation*}
The equality holds whenever $1_s-\tau_{i,s}-\sigma_{i,s}-\gamma_s$ for $i=1,2.$

In summary, the leading order of terms in the above sum is $O\left (d^{-2\#(S_1)-3\#(S_2)-3\#(S_3)} \right)$. Hence the leading order of the variance is $O(d^{-2}),$ which leads to the almost sure convergence. 
\end{proof}


\subsection{Empirical eigenvalue distribution of $H$}\label{subsec:main}

Let $(\mathcal{A}, \phi)$ be a noncommutative probability space and $a_1, \ldots, a_{m} $ be a sequence of $\epsilon$-freely independent random variables in $(\mathcal{A}, \phi)$ such that the law of $a_\iota$
is $\mu_\iota$ for each $\iota=1,\ldots, m.$ Because the local terms $H_1, \ldots, H_m$ are asymptotic $\epsilon$-free, the sum $H = \sum_{\iota=1}^m H_\iota$ converges in moment to the sum $\sum_{\iota=1}^{m} a_\iota$ \cite{CC2015}[Theorem 4]. We will provide an alternate combinatoric proof which will use the moment-cumulant formulas \eqref{eq:moment-cumulant} and \eqref{eq:inversion-moment-cumulant}. Furthermore, due to the results in Subsection \ref{subsec:almost-surely}, we have an almost sure convergence for $H$.

\begin{thm}(Restatement of Theorem \ref{thm:GUE}).
As $d \rightarrow \infty,$ $\mu_H$ converges weakly, almost surely, to $\sqsum^{(m)}_{\epsilon}\mu_{sc}$, if $H_{\iota}$'s are independently sampled by GUE. 
\end{thm}

\begin{proof}
Let $(\mathcal{A}, \phi)$ be a noncommutative probability space and $s_1, \ldots, s_m $ be a sequence of $\epsilon$-freely independent random variables in $(\mathcal{A}, \phi)$ which satisfies the semicircular law $\mu_{sc}$. For any even $k\geq 1,$ the $k$-th moment of $\sum_{\iota=1}^{m}s_\iota$ is given by
\begin{equation}\label{eq:almost-GUE}
\begin{split}
\phi \left( \sum_{\iota=1}^{m} s_\iota \right)^k &= \sum_{\iota_1, \ldots, \iota_k=1}^m \phi( s_{\iota_1} s_{\iota_2} \cdots s_{\iota_k})\\
& = \sum_{{\boldsymbol{\iota}}: [k] \rightarrow [m]}\sum_{\alpha \in \mathcal{NC}^{(\epsilon, {\boldsymbol{\iota}})}(k)} \kappa_\alpha^{(\epsilon)} [s_{\iota_1}, s_{\iota_2}, \ldots, s_{\iota_k}]\\
& = \sum_{{\boldsymbol{\iota}}: [k] \rightarrow [m]}\sum_{\alpha \in \mathcal{NC}^{(\epsilon, {\boldsymbol{\iota}})}(k)} \kappa_\alpha [s_{\iota_1}, s_{\iota_2}, \ldots, s_{\iota_k}]\\
& =   \sum_{{\boldsymbol{\iota}}: [k]\rightarrow [m]} \sum_{\alpha \in  \mathcal{NC}_2^{(\epsilon, {\boldsymbol{\iota}})}(k)} 1,
\end{split}   
\end{equation}
where we have used the following fact about the free cumulant of semicircular elements  \cite{NS2006}:
\begin{equation*}
\kappa_\alpha [s_{\iota_1}, s_{\iota_2}, \ldots, s_{\iota_k}]=  \begin{cases} 1 & \text{if} \; \alpha \;\text{is a pairing};\\
0 &  \text{otherwise}.
\end{cases}
\end{equation*}

Moreover, note that for odd $k,$ $\phi \left( \sum_{\iota=1}^m s_\iota \right)^k =0.$ Hence by combining Proposition \ref{prop:moment-GUE} and \ref{prop:almost-GUE}, almost surely we have 
\begin{equation*}
\begin{split}
\lim_{d \rightarrow \infty} {\rm tr} H^k & = \lim_{d \rightarrow \infty} \sum_{{\boldsymbol \iota}:[k] \rightarrow [m]} {\rm tr} \left( H_{\iota_1} \cdots H_{\iota_k} \right)\\
&= \sum_{{\boldsymbol{\iota}}: [k]\rightarrow [m]} \sum_{\pi \in  \mathcal{NC}_2^{(\epsilon, {\boldsymbol{\iota}})}(k)} 1\\
&= \phi \left( \sum_{\iota=1}^m s_\iota \right)^k.
\end{split}
\end{equation*}
It follows that $\mu_H$ converges weakly, almost surely to $\sqsum^{(m)}_{\epsilon}\mu_{sc}$ as $d \rightarrow \infty.$
\end{proof}


\begin{thm}(Restatement of Theorem \ref{thm:U}).
As $d \rightarrow \infty,$  $\mu_H$ converges weakly, almost surely, to $\mu_1 \sqsum_{\epsilon} \cdots \sqsum_{\epsilon} \mu_m$, if $H_{\iota}$'s are independently sampled by the Haar unitary invariant ensemble which obeys Assumption \ref{ass:1}. 
\end{thm}

\begin{proof}
Let $(\mathcal{A}, \phi)$ be a noncommutative probability space and $a_1, \ldots, a_{m} $ be a sequence of $\epsilon$-freely independent random variables in $(\mathcal{A}, \phi)$ such that the law of $a_\iota$
is $\mu_\iota$ for each $\iota=1,\ldots, m$. By Assumption \ref{ass:1}, for every $\alpha \in \mathcal{P}(k)$ and ${\boldsymbol \iota}: [k] \rightarrow [m]$  we have 
$$\lim_{d\rightarrow \infty}{\rm tr}_{\sigma} \left[A_{\iota_1}, \ldots, A_{\iota_k} \right] =  \phi_\alpha \left[ a_{\iota_1},\ldots, a_{\iota_k} \right],$$
where $\sigma = P_\alpha.$

Then for any $k\geq 1,$ the $k$-th moment of $\sum_{\iota=1}^{m}a_\iota$ is given by
\begin{equation*}
\begin{split}
\phi \left( \sum_{\iota=1}^{m} a_\iota \right)^k &= \sum_{\iota_1, \ldots, \iota_k=1}^{m} \phi( a_{\iota_1} a_{\iota_2} \cdots a_{\iota_k})\\
& =   \sum_{{\boldsymbol{\iota}}: [k]\rightarrow [m]}  \sum_{\alpha\in \mathcal{NC}^{(\epsilon,{\boldsymbol{\iota}})}(k)} \kappa_\alpha^{(\epsilon)} [a_{\iota_1}, \ldots, a_{\iota_k}]\\
& = \sum_{{\boldsymbol{\iota}}: [k] \rightarrow [m]} \sum_{\substack{\alpha,\beta \in \mathcal{NC}^{(\epsilon, {\boldsymbol{\iota}})}(k), \\ \beta \leq \alpha}}\phi_\beta [a_{\iota_1}, \ldots, a_{\iota_k}] \cdot \mu_{\mathcal{NC}^{(\epsilon, {\boldsymbol{\iota}})}(k)} (\beta, \alpha).
\end{split}   
\end{equation*}
By Theorem \ref{thm:iso} and Proposition \ref{prop:mobius}, it follows that
\begin{equation*}
\begin{split}
\phi \left( \sum_{\iota=1}^{m} a_\iota \right)^k & =\sum_{{\boldsymbol{\iota}}: [k] \rightarrow [m]}  \sum_{\substack{\sigma, \tau \in S^{(\epsilon,{\boldsymbol{\iota}})}_{NC}(\gamma);\\ \tau \leq \sigma }} \mu(\sigma \tau^{-1}) \cdot \lim_{d\rightarrow \infty} {\rm tr}_{\tau}\left[ A_{\iota_1}, \ldots, A_{\iota_k}\right],
\end{split}   
\end{equation*}
where $\sigma = P_\alpha$ and $\tau = P_\beta.$ Note we have used the fact that
\begin{equation*}
\begin{split}
\mu(\sigma \tau^{-1}) & = \prod_{B \in \ker{\boldsymbol \iota}} \mu(\sigma_B\tau_B^{-1})=  \prod_{B \in \ker{\boldsymbol \iota}} \mu_{\mathcal{NC}(\#(B))}(\beta_B, \alpha_B)\\
& =  \mu_{\mathcal{NC}^{(\epsilon, {\boldsymbol{\iota}})}(k)} (\beta, \alpha).
\end{split}
\end{equation*}

Hence by combining Proposition \ref{prop:moment-U} and \ref{prop:almost-U}, almost surely we have 
\begin{equation*}
\begin{split}
\lim_{d \rightarrow \infty} {\rm tr} H^k & = \lim_{d \rightarrow \infty} \sum_{{\boldsymbol \iota}:[k] \rightarrow [m]} {\rm tr} \left( H_{\iota_1} \cdots H_{\iota_k} \right)\\
&= \sum_{{\boldsymbol{\iota}}: [k]\rightarrow [m]}  \sum_{\substack{\sigma, \tau \in S^{(\epsilon,{\boldsymbol{\iota}})}_{NC}(\gamma);\\ \tau \leq \sigma }} \mu(\sigma \tau^{-1}) \cdot \lim_{d\rightarrow \infty} {\rm tr}_{\tau}\left[ A_{\iota_1}, \ldots, A_{\iota_k}\right]\\
&= \phi \left( \sum_{\iota=1}^m s_\iota \right)^k.
\end{split}
\end{equation*}
It follows that $\mu_H$ converges weakly, almost surely to $\mu_1 \sqsum_{\epsilon} \cdots \sqsum_{\epsilon} \mu_m$ as $d \rightarrow \infty.$
\end{proof}


\section{Bounds for the largest eigenvalue}

Let  $\lambda_{\max}$ be the largest eigenvalue of $H,$ we have the following bounds for $\lambda_{\max}.$ 

\begin{prop}\label{prop:lowerH-GUE}
If $H_\iota$'s are independently sampled by GUE, then with probability one, we have
\begin{equation}
\lambda_{\max} \geq 2\sqrt{m}, \; \text{as} \; d \rightarrow \infty.
\end{equation}
\end{prop}

\begin{proof}
Suppose that $\ker{\boldsymbol \iota}$ is noncrossing, then by the definition of $\mathcal{NC}^{(\epsilon, {\boldsymbol \iota})}(k)$ we have 
\begin{equation*}
\begin{split}
 \mathcal{NC}^{(\epsilon, {\boldsymbol \iota})}(k) =  \{\alpha \in  \mathcal{NC}(k): \alpha \leq \ker{\boldsymbol \iota}\}.
\end{split}
\end{equation*}

By Proposition \ref{prop:almost-GUE} almost surely we have
\begin{equation*}
\begin{split}
\lim_{d \rightarrow \infty} {\rm tr} \left( H^{2k}\right) & = \sum_{{\boldsymbol \iota}: [2k] \rightarrow [m]}\sum_{\alpha \in  \mathcal{NC}_2^{(\epsilon,{\boldsymbol{\iota}})}(2k)} 1\\
& = \sum_{\beta \in \mathcal{P}(2k)} \sum_{\substack{{\boldsymbol \iota}: [2k] \rightarrow [m],\\ \ker{\boldsymbol \iota}= \beta}} \sum_{\alpha \in  \mathcal{NC}_2^{(\epsilon,{\boldsymbol{\iota}})}(2k)}1\\
& \geq \sum_{\beta \in \mathcal{NC}(2k)} \sum_{\substack{{\boldsymbol \iota}: [2k] \rightarrow [m],\\ \ker{\boldsymbol \iota}= \beta}} \sum_{\substack{\alpha \in  \mathcal{NC}_2 (2k),\\ \alpha \leq \ker{\boldsymbol \iota}}} 1\\
& =  \sum_{\beta \in \mathcal{NC}(2k)} \sum_{\substack{\alpha \in  \mathcal{NC}_2 (2k),\\ \alpha \leq \beta}} \sum_{\substack{{\boldsymbol \iota}: [2k] \rightarrow [m],\\ \ker{\boldsymbol \iota}= \beta}}  1\\
& \geq  \sum_{\beta \in \mathcal{NC}_2(2k)} \sum_{\substack{{\boldsymbol \iota}: [2k] \rightarrow [m],\\ \ker{\boldsymbol \iota}= \beta}}  1\\
& \simeq m^k \cdot C_{k},
\end{split}
\end{equation*}
where $C_k = \frac{1}{k+1} \binom{2k}{k}$ is the Catalan number. Note that we have by Stirling's formula
\begin{equation}
\lim_{k \rightarrow \infty} C_k^{\frac{1}{k}} = 4.
\end{equation}

Then for any $\delta >0,$ and sufficient large $d$
\begin{equation*}
\begin{split}
{\rm \mathbb{P}}[\lambda_{\max} < 2\sqrt{m}- \delta] & \leq {\rm \mathbb{P}} [ {\rm tr} \left( H^{2k}\right) < (2\sqrt{m}-\delta)^{2k}]\\
& \leq {\rm \mathbb{P}} [ \lim_{k \rightarrow \infty}\left({\rm tr} \left( H^{2k}\right) \right)^{\frac{1}{2k}}\leq 2\sqrt{m}-\delta]\\
& = {\rm \mathbb{P}} [ 2\sqrt{m} \leq 2\sqrt{m}-\delta]=0.
\end{split}
\end{equation*}
Hence ${\rm \mathbb{P}}[\lambda_{\max} < 2\sqrt{m}- \delta] \rightarrow 0$ as $d\rightarrow \infty,$  which completes our proof.
\end{proof}

\begin{prop}
Suppose that $H_{\iota}$'s are independently sampled by a Haar unitary invariant ensemble which obeys Assumption \ref{ass:1}, and for each $\iota,$ $\mu_\iota$ is distributed with mean 
$0$ and variance $1,$ then almost surely we have 
\begin{equation}
\lambda_{\max} \geq 2\sqrt{m}, \; \text{as} \; d \rightarrow \infty.
\end{equation}
\end{prop}

\begin{proof}
Let $(\mathcal{A}, \phi)$ be a noncommutative probability space and $a_1, \ldots, a_{m} $ be a sequence of $\epsilon$-freely independent random variables in $(\mathcal{A}, \phi)$ such that the law of $a_\iota$
is $\mu_\iota$ for each $\iota=1,\ldots, m$. By our assuption, for every $\alpha \in \mathcal{P}(k)$ and ${\boldsymbol \iota}: [k] \rightarrow [m]$  we have 
$$\lim_{d\rightarrow \infty}{\rm tr}_{\sigma} \left[A_{\iota_1}, \ldots, A_{\iota_k} \right] =  \phi_\alpha \left[ a_{\iota_1},\ldots, a_{\iota_k} \right],$$
where $\sigma = P_\alpha.$ Moreover, $\phi(a_\iota)=0$ and $\phi(a_\iota^2)=1$ for each $\iota=1, \ldots, m.$

By Proposition \ref{prop:almost-U}, almost surely we have
\begin{equation}
\begin{split}
\lim_{d \rightarrow \infty} {\rm tr} \left(H^{2k}\right) & = \sum_{{\boldsymbol{\iota}}: [2k]\rightarrow [m]}  \sum_{\alpha\in \mathcal{NC}^{(\epsilon,{\boldsymbol{\iota}})}(2k)} \kappa_\alpha^{(\epsilon)} [a_{\iota_1}, \ldots, a_{\iota_{2k}}]\\
& \geq \sum_{{\boldsymbol{\iota}}: [2k]\rightarrow [m]}  \sum_{\substack{\alpha \in  \mathcal{NC} (2k),\\ \alpha \leq \ker{\boldsymbol \iota}}} \kappa_\alpha [a_{\iota_1}, \ldots, a_{\iota_{2k}}]\\
& = \sum_{\beta \in \mathcal{NC}(2k)} \sum_{\substack{\alpha \in  \mathcal{NC} (2k),\\ \alpha \leq \beta}} \sum_{\substack{{\boldsymbol \iota}: [2k] \rightarrow [m],\\ \ker{\boldsymbol \iota}= \beta}}  \kappa_\alpha [a_{\iota_1}, \ldots, a_{\iota_{2k}}]\\
& \geq \sum_{\beta \in \mathcal{NC}_2(2k)} \sum_{\substack{{\boldsymbol \iota}: [2k] \rightarrow [m],\\ \ker{\boldsymbol \iota}= \beta}}  \kappa_\beta [a_{\iota_1}, \ldots, a_{\iota_{2k}}]\\
& \geq  \sum_{\beta \in \mathcal{NC}_2(2k)} \sum_{\substack{{\boldsymbol \iota}: [2k] \rightarrow [m],\\ \ker{\boldsymbol \iota}= \beta}}  1\\
& \simeq m^k \cdot C_{k},
\end{split}
\end{equation}
where we have used the fact that $\kappa_\alpha^{(\epsilon)} [a_{\iota_1}, \ldots, a_{\iota_{2k}}] = \kappa_\alpha [a_{\iota_1}, \ldots, a_{\iota_{2k}}]$ for $\alpha \in  \mathcal{NC} (2k),$ and
$$ \kappa_\beta [a_{\iota_1}, \ldots, a_{\iota_{2k}}] =1$$
for $\beta \in \mathcal{NC}_2(2k)$ and $\ker{\boldsymbol \iota} = \beta.$
The rest of the proof is the same as the proof of Proposition \ref{prop:lowerH-GUE}.
\end{proof}

Remark that the above lower bound is universal for all $\epsilon$ and it becomes optimal for the extreme case whenever every non-diagonal entry of $\epsilon$ is $0.$ For example, let $\mathcal{K}_\iota = [n]$ for every $\iota \in [m]$, then almost surely we have \cite{H78,CM14,HT05}
$$\lambda_{\max} \leq 2 \sqrt{m}, \;\; \text{as} \; d \rightarrow \infty.$$

On the other hand, there is a trivial upper bound for $\lambda_{\max}$ without considering the interactions between $H_{\iota}$'s. For simplicity, we only consider the case of $H_\iota$ being GUE. Note that it is  well-known \cite{BY88}  that $\|H_\iota\|_\infty =2$ almost surely as $d \rightarrow \infty,$ then we have 
\begin{equation}\label{eq:tri-upper-bound}
\begin{split}
\lambda_{\max} & \leq \|H\|_\infty \leq \sum_{\iota=1}^m \|H_\iota\|_\infty\\
& \leq 2m,
\end{split}
\end{equation}
as $d \rightarrow \infty$. The above upper bound can be improved by considering the interactions. 

Now let us consider the following XY-model: Let $m=n-1$ and assume that 
\begin{equation}\label{eq:XY-model}
\mathcal{K}_\iota = \{\iota, \iota+1\}, \iota=1, \ldots, n-1.
\end{equation}
Thus the entries of $\epsilon$ are given by ($i \geq j$)
\begin{equation*}
\epsilon_{i,j}= \begin{cases} 0 & \text{if}\;\; i=j, j+1;\\
1 &  \text{others}.
\end{cases}
\end{equation*}
Moreover, for given ${\boldsymbol \iota}: [k] \rightarrow [n-1]$ we have $J_1 = \{j \in [k]: \iota_j =1\},$ $J_s = \{j \in [k]: \iota_j = s-1, s\}, s=2, \ldots, n-1,$ and $J_{n}= \{j \in [k]: \iota_j =n-1\}.$ 

\begin{prop}\label{prop:XY-model}
Let $m =n-1,$ and consider the Hamiltonian $H$ with the interactions given by \eqref{eq:XY-model}, if $H_\iota$'s are independently sampled by GUE, then with probability one, we have
\begin{equation}\label{eq:upper-bound}
\lambda_{\max} \leq 2(n-3)+ 2\sqrt{2}, \;\; \text{as} \; d\rightarrow \infty.
\end{equation}
\end{prop}

\begin{proof}
For given ${\boldsymbol \iota}: [k] \rightarrow [n-1]$, by the definition of $\mathcal{NC}^{(\epsilon, {\boldsymbol \iota})}(k)$ we have the following iteration relation
\begin{equation*}
\begin{split}
 \mathcal{NC}^{(\epsilon, {\boldsymbol \iota})}(k) & \subseteq \mathcal{NC}(J_1) \times \mathcal{NC}^{(\epsilon, {\boldsymbol \iota}_1)}([k]/J_1),
\end{split}
\end{equation*}
where ${\boldsymbol \iota}_1$ is the restriction of ${\boldsymbol \iota}$ on $[k]/ J_1.$

Starting with Equation \eqref{eq:moment-GUE} we have
\begin{equation*}
\begin{split}
\mathbb{E} \cdot {\rm tr} \left( H \right)^{k} & = \sum_{{\boldsymbol \iota}: [k] \rightarrow [n-1]}\sum_{\alpha \in  \mathcal{NC}_2^{(\epsilon,{\boldsymbol{\iota}})}(k)} 1 + O\left(\frac{1}{d} \right)\\
& \leq \sum_{{\boldsymbol \iota}: [k] \rightarrow [n-1]}\sum_{\alpha_1 \in \mathcal{NC}_2(J_1)}\sum_{\alpha_2 \in  \mathcal{NC}_2^{(\epsilon,{\boldsymbol{\iota}}_1)}([k]/J_1)} 1 +O\left(\frac{1}{d} \right)\\
& = \sum_{\ell_1=0}^{k} \binom{k}{\ell_1} \# (\mathcal{NC}_2(\ell_1)) \cdot \sum_{{\boldsymbol \iota}: [k-\ell_1] \rightarrow [2,n-1]}\sum_{\alpha_2 \in  \mathcal{NC}_2^{(\epsilon,{\boldsymbol{\iota}})}(k-\ell_1)} 1+ O\left(\frac{1}{d} \right)\\
& =  \sum_{\ell_1=0}^{k} \binom{k}{\ell_1} \# (\mathcal{NC}_2(\ell_1)) \cdot \mathbb{E} {\rm tr} \left( H_2+ \cdots + H_{n-1}\right)^{k-\ell}+ O\left(\frac{1}{d} \right).
\end{split}
\end{equation*}
By an inductive argument, we have
\begin{equation*}
\begin{split}
\mathbb{E} \cdot {\rm tr} \left( H \right)^{k} & \leq  \sum_{\ell_1=0}^{k} \binom{k}{\ell_1} \# (\mathcal{NC}_2(\ell_1)) \cdot \mathbb{E} {\rm tr} \left( H_2+ \cdots + H_{n-1}\right)^{k-\ell_1}+ O\left(\frac{1}{d} \right)\\
& \leq \sum_{\ell_1=0}^{k}\sum_{\ell_2=0}^{k-\ell_1}   \binom{k}{\ell_1} \binom{k-\ell_1}{\ell_2} \# (\mathcal{NC}_2(\ell_1)) \cdot \# (\mathcal{NC}_2(\ell_2)) \cdot \mathbb{E} {\rm tr} \left( H_3+ \cdots + H_{n-1}\right)^{k-\ell}+ O\left(\frac{1}{d} \right)\\
& \ldots \; \ldots\\
& \leq \sum_{\ell_1=0}^{k}\sum_{\ell_2=0}^{k-\ell_1} \cdots \sum_{\ell_{n-3}=0}^{k-(\ell_1+\cdots +\ell_{n-4})} \binom{k}{\ell_1} \binom{k-\ell_1}{\ell_2}\cdots \binom{k-(\ell_1+\cdots+\ell_{n-4})}{\ell_{n-3}}
\prod_{i=1}^{n-3} \# (\mathcal{NC}_2(\ell_i))\times\\
& \times \mathbb{E} {\rm tr} \left( H_{n-2} + H_{n-1}\right)^{k-(\ell_1+\cdots +\ell_{n-3})}+ O\left(\frac{1}{d} \right)\\
& = \sum_{\ell_1=0}^{k}\sum_{\ell_2=0}^{k-\ell_1} \cdots \sum_{\ell_{n-3}=0}^{k-(\ell_1+\cdots +\ell_{n-4})} \binom{k}{\ell_1} \binom{k-\ell_1}{\ell_2}\cdots \binom{k-(\ell_1+\cdots+\ell_{n-4})}{\ell_{n-3}}
\prod_{i=1}^{n-3} \# (\mathcal{NC}_2(\ell_i))\times\\
& \times  \# (\mathcal{NC}_2(k-(\ell_1+\cdots+\ell_{n-3})) \cdot 2^{\frac{k-(\ell_1+\cdots +\ell_{n-3})}{2}}+ O\left(\frac{1}{d} \right)\\
& \leq \sum_{\ell_1=0}^{k}\sum_{\ell_2=0}^{k-\ell_1} \cdots \sum_{\ell_{n-3}=0}^{k-(\ell_1+\cdots +\ell_{n-4})} \binom{k}{\ell_1} \binom{k-\ell_1}{\ell_2}\cdots \binom{k-(\ell_1+\cdots+\ell_{n-4})}{\ell_{n-3}}
4^{\frac{k}{2}} \cdot 2^{\frac{k-(\ell_1+\cdots +\ell_{n-3})}{2}}+ O\left(\frac{1}{d} \right)\\
& := M_k +  O\left(\frac{1}{d} \right),
\end{split}
\end{equation*}
where we have used the fact that $\# (\mathcal{NC}_2(k))= C_{k/2} \leq 4^{k/2}.$ Note that we have by Stirling's formula
\begin{equation}
\lim_{k \rightarrow \infty} M_k^{\frac{1}{k}} = 2(n-3)+ 2\sqrt{2}.
\end{equation}

Choose a sequence $k(d) \rightarrow \infty$ as $d \rightarrow \infty$ such that 
$$M_{k(d)} \leq \left( 2(n-3)+ 2\sqrt{2} \right)^{k(d)},\; \text{and}\;  k(d)/\log d \rightarrow 0\; \text{as} \; d \rightarrow \infty.$$
Then for any $\delta >0,$ and sufficient large $d,$
\begin{equation*}
\begin{split}
{\rm \mathbb{P}}[\lambda_{\max} >2(n-3)+ 2\sqrt{2}+ \delta] & \leq {\rm \mathbb{P}} [ {\rm Tr} \left( H^{2k(d)}\right) > (2(n-3)+ 2\sqrt{2}+ \delta)^{2k(d)}]\\
& \leq \frac{ \mathbb{E} \cdot{\rm Tr} \left( H^{2k(d)}\right)}{(2(n-3)+ 2\sqrt{2}+ \delta)^{2k(d)}}\\
& \leq \frac{ d^n \cdot (2(n-3)+ 2\sqrt{2})^{2k(d)}}{(2(n-3)+ 2\sqrt{2}+ \delta)^{2k(d)}}.
\end{split}
\end{equation*}
Hence ${\rm \mathbb{P}}[\lambda_{\max} >2(n-3)+ 2\sqrt{2}+ \delta] \rightarrow 0$ as $d\rightarrow \infty,$  which completes our proof.
\end{proof}


\section{Conclusion and discussion}

In this paper, we consider the spectrum of local Hamiltonians given by some random ensembles in the large $d$ limit. Since the Hamiltonian is local, the correlations between the local terms are somewhat complicated. However, as pointed out in \cite{ML19} and \cite{CC2015}, the local terms of the Hamiltonian are asymptotically $\epsilon$-free (or equivalently heap-free) in the limit, namely, it is appropriate to carry out our investigation in the framework of noncommutative probability theory. Therefore, it is natural to say that the spectrum distribution of the total Hamiltonian is some convolution of its local terms' probability distributions. Our work has two main contributions: (i) by proper variance estimations, the asymptotic $\epsilon$-freeness of the local terms holds generically. As a byproduct, our proof may shed some light on considering the second order $\epsilon$-freeness of the corresponding random matrix models. Such a problem is inspired by the work of second order freeness \cite{2004Annular, MSS2007, 2006Second, 2012Second}.
Moreover, the almost sure convergence of the matrix model may have further applications in operator algebra. (ii) the isomorphism between the $\epsilon$-noncrossing partitions and permutations brings us a new bridge to connect the matrix model and its large $d$ limit, just like what Biane had done in the free probability theory. And it allows us to provide a direct combinatoric proof for the convergence of the total Hamiltonian $H$, though the convergence in moments can be induced by the results in \cite{CC2015}. Since the enumeration of the $\epsilon$-noncrossing partitions/permutations is unclear; our results are only formally meaningful. Nevertheless, we can derive bounds for the largest eigenvalue of the total Hamiltonian via some rough estimation of the enumeration. We end our conclusion with the following questions.
\begin{enumerate}[{\rm (i)}]  
\item Is it possible to explicitly calculate the $\epsilon$-free convolution for some given non-trivial $\epsilon$? For the GUE case, the expectation of $k$th-moment of $H$ is given by 
$$\mathbb{E} \cdot {\rm tr} (H^k)=\sum_{{\boldsymbol \iota}: [k] \rightarrow [m]} \sum_{\alpha \in \mathcal{NC}_2^{(\epsilon, {\boldsymbol \iota})}(k)} 1.$$ 
Can one connect the above combinatorial summation to some transformations of $\mu_{sc}$, e.g., the Cauchy transform of $\mu_{sc}$? 
\item Assume that $H_{\iota}$ is given by independent projection, i.e., $H_\iota =   U_\iota P_\iota U_\iota^* \otimes \bigotimes_{s \notin \mathcal{K}_\iota} (\mathbb{C} \un_d),$ where $P_\iota$ is a rank $p$ projection. Then $\mu_\iota$ converges to the Bernoulli distribution for each $\iota.$ Is it possible to figure out the spectrum gap of $H$ in the thermal dynamical limit ($m \rightarrow \infty$)? The XY-model given by \eqref{eq:XY-model}
is of great interest. Since it can be shown that the Hamiltonian of this model is generically gapless \cite{M17}. There is some evidence that it is possible to recover this conclusion by estimating the density of $\epsilon$-free convolution. Consider the extreme case when $\mathcal{K}_\iota=[n],$ the density of $\mu_H$ becomes continuous when $m \rightarrow \infty$ \cite{BV95, Huang15}. Is it possible to make any similar qualitative estimation for other non-trivial interactions (hence for some non-trivial $\epsilon$)? 
\item In \cite{BML19}, the authors considered the dynamical correlator under the Hamiltonian to be given by the sum of two independent random matrices. So it is natural to consider the correlator under the local random Hamiltonians. Moreover, it is interesting to consider the convergence to equilibrium under a local random Hamiltonian. For the global case, see \cite{BCHHKM12}.
\end{enumerate}

\vspace{2mm}

{\it Acknowledgments}--We would like to thank Ian Charlesworth, Akihiro Miyagawa, and Simeng Wang for valuable remarks and fruitful discussions. ZY would like to thank Michal Horodecki and Michal Studzinski for enlightening discussions about local 
random Hamiltonians. LZ and ZY are partially supported by NSFC No. 12031004.


\appendix

\section{Some notions of combinatorics}\label{sec:preli} 

In this appendix, we will recall some combinatoric notations and facts frequently appearing in this paper, which mainly come from  \cite{NS2006, MS17} and the references therein.    

For natural numbers $m, n \in \mathbb{N}$ with $m<n,$ we denote by $[m, n]$ the interval of natural numbers between $m$ and $n,$
$$[m,n]: = \{ m, m+1, \ldots, n-1, n\}.$$
Moreover, we use $[n]$ to denote $[1, n].$ 

\vspace{2mm}

\noindent{\it Permutations}--We denote the set of permutations on $n$ elements by $S_n$. For a permutation $\pi \in S_n$ we denote $\# (\pi)$ by the number of cycles of $\pi$ and by $|\pi|$
the minimal number of transpositions needed to decompose $\pi.$ There is a nice relation between $\#(\pi)$ and $|\pi|$, namely,
\begin{equation} 
|\pi| + \# (\pi) = n, \;\; \text{for all} \; \pi \in S_n.
\end{equation}

Let $\gamma_n \in S_n$ be the full cycle $\gamma_n = (1,2, \ldots, n).$ For any $\pi \in S_n$ we always have 
$$|\pi|+ |\pi^{-1} \gamma_n| \geq n-1.$$
We call $\pi$ noncrossing if the equality holds, or equivalently we can say that $\pi$ lies on the geodesic path $1_n-\pi-\gamma_n$ in $S_n$. We denote the set of all noncrossing permutations in $S_n$ by $S_{NC}(\gamma_n).$ For $\sigma, \pi \in S_{NC}(\gamma_n).$ We say that $\sigma \leq \pi$ if $\sigma$ and $\pi$ lie on the same geodesic path and $\sigma$ comes before $\pi,$ i.e., $1_n-\sigma-\pi-\gamma_n.$ The set $S_{NC}(\gamma_n)$ endowed with $"\leq"$ becomes a poset.  

Fix $m, n \in \mathbb{N}$ and denote by $\gamma_{m,n}$ the product of the two cycles
\begin{equation*}
\gamma_{m,n} = (1, 2, \ldots, m) \cdot (m+1, m+2, \ldots, m+n).
\end{equation*}
A permutation $\pi \in S_{m+n}$ is called connected if the pair $\pi$ and $\gamma_{m,n}$ generates a transitive subgroup in $S_{m+n}.$ We note that a connected permutation $\pi \in S_{m+n}$ always satisfies 
\begin{equation}
m+n \leq |\pi|+ |\pi^{-1}\gamma_{m,n}|.
\end{equation}
If $\pi \in S_{m+n}$ is connected, and if we have equality, then we call $\pi$ annular noncrossing. More general, suppose $\pi \in S_{m+n}$ is connected, then we have the following formula
\begin{equation}\label{eq:Euler}
\# (\pi) + \# (\pi^{-1} \gamma_{m,n}) + \# (\gamma_{m,n} )= m+n +2(1-g),
\end{equation}  
where $g\geq 0$ is the genus of $\pi$ relative to $\gamma_{m,n}.$ 

\vspace{2mm}

\noindent{\it Partitions}--We say $\alpha=\{A_1, A_2, \ldots, A_k\}$ is a partition of $[n]$ if the sets $A_i$ are disjoint and non-empty and their union is equal to $[n].$ We call $A_1, \ldots, A_k$ the blocks of partition $\alpha.$ The set of all partitions of $[n]$ is denoted by $\mathcal{P}(n).$ If there are only two elements in every block of $\alpha \in \mathcal{P}(n)$, we call $\alpha$ a pair partition and the set of all pair partitions of $[n]$ is denoted by
$\mathcal{P}_2(n).$  
If $\alpha=\{A_1, \ldots, A_k\}$ and $\beta= \{B_1, \ldots, B_s\}$ are partitions of $[n],$ we say that $\alpha \leq \beta$ if for every block $A_i$ there exists some block $B_j$ such that $A_i \subseteq B_j.$ With the relation $"\leq"$, the set $\mathcal{P}(n)$ is also a poset. 

Given two elements $i, j \in [n],$ we write $i \sim_\alpha j$ if $i$ and $j$ belongs to the same block of $\alpha. $ A partition $\alpha$ is called a crossing if $i_1 < j_1 < i_2 < j_2 \in [n]$ exist such that $i_1 \sim_\alpha i_2 \nsim_\alpha j_1 \sim_\alpha j_2.$ We call $\alpha$ a noncrossing partition if $\alpha$ is not crossing, and we denote $NC(n)$ (resp. $NC_2(n)$) by the set of all noncrossing partitions (resp. noncrossing pair partitions) of $[n].$ It is known that the cardinality of the set of noncrossing partitions is the Catalan number, i.e.,
\begin{equation}
\#(\mathcal{NC}(k)) = \# (\mathcal{NC}_2(2k)) = C_k, 
\end{equation}
where $C_k = \frac{1}{k+1} \binom{2k}{k}.$

\vspace{2mm}

\noindent{\it Permutations vs. Partitions}--A permutation can always be written as a product of cycles, so one can identify a permutation $\pi \in S_n$ with a partition $\alpha \in \mathcal{P}(n)$ by omitting the order on the cycles. 
In particular, there is a one-to-one identification between the transpositions and the pair partitions. We remark that now $n$ should be even. In this sense, we can multiply a pair partition $\alpha \in \mathcal{P}_2(n)$ with a permutation $\pi \in S_n,$ and their product $\alpha \cdot \pi$ is viewed as an element in $S_n.$ Moreover, we always have 
\begin{equation}
\# (\alpha \cdot \gamma_n) \leq \frac{n}{2} + 1,
\end{equation}
the equality holds whenever $\alpha$ is noncrossing.  

For the noncrossing case, the situation is much better; we recall the following fact due to Biane \cite{B97}:
\begin{prop}\label{prop:Biane}
There is a bijection between $NC(n)$ and the set $S_{NC}(\gamma_n)$, which preserves the poset structure. 
\end{prop}


\bibliography{reference}

\end{document}